\documentclass[preprint,authoryear,twocolumn,3p]{elsarticle}

\journal{Environmental Modeling and Software}

\usepackage{graphicx}
\usepackage{subfig}
\usepackage{caption}
\usepackage{amssymb}
\usepackage{amsmath}
\usepackage{natbib}
\usepackage{tikz}
\usetikzlibrary{arrows,backgrounds,positioning}
\usetikzlibrary{fit}					
\usetikzlibrary{backgrounds}	
\usepackage{pgf}
\usepackage[bookmarksnumbered = true, colorlinks = true, linktocpage = true, bookmarksopen = true,
    citecolor=blue,%
    filecolor=blue,%
    linkcolor=blue,%
    urlcolor=blue,
]{hyperref}
\renewcommand{\vec}[1]{\mbox{\boldmath{$#1$}}}
\newcommand{\mat}[1]{\mathrm{#1}}

\newcommand{\pipe}{\;\middle\vert\;}

\usepackage{rotating}

\usepackage{fancyvrb}
\usepackage{booktabs}
\usepackage{tabularx}

\usepackage{tikz}
\usetikzlibrary{arrows,positioning} 
\tikzset{
    >=stealth',
    punkt/.style={
           rectangle,
           rounded corners,
           draw=black, very thick,
           minimum height=2em,
           minimum width=6.5em,
           text centered},
    pil/.style={
           ->,
           thick,
           shorten <=2pt,
           shorten >=2pt,}
}

\date{}

\begin{document}
\begin{frontmatter}

\title{Bayesian semi-parametric forecasting of ultrafine particle number concentration with penalised splines and autoregressive errors}

\author[cpme]{Sam Clifford\corref{cor1}}\ead{sj.clifford@student.qut.edu.au}
\author[uhp]{Bjarke M{\o}lgaard}
\author[qut]{Sama {Low Choy}}
\author[uhm]{Jukka Corander}
\author[uhp]{Kaarle H{\"a}meri}
\author[qut]{Kerrie Mengersen}
\author[uhp,uj]{Tareq Hussein}
\address[cpme]{School of Chemistry, Physics and Mechanical Engineering, Science and Engineering Faculty, Queensland University of Technology, GPO Box 2434, Brisbane, Qld 4000, Australia}
\address[uhp]{Department of Physics, University of Helsinki, P.O. Box 48, FI-00014, Finland}
\address[qut]{School of Mathematical Sciences, Science and Engineering Faculty, Queensland University of Technology, GPO Box 2434, Brisbane, Qld 4000, Australia}
\address[uhm]{Department of Mathematics and Statistics, University of Helsinki, P.O. Box 68, FI-00014, Finland}
\address[uj]{Department of Physics, University of Jordan, Amman, 11942, Jordan}

\cortext[cor1]{Corresponding author}

\begin{abstract}
Observational time series data often exhibit both cyclic temporal trends and autocorrelation and may also depend on covariates. As such, there is a need for flexible regression models that are able to capture these trends and model any residual autocorrelation simultaneously. Modelling the autocorrelation in the residuals leads to more realistic forecasts than an assumption of independence. In this paper we propose a method which combines spline-based semi-parametric regression modelling with the modelling of auto-regressive errors.

The method is applied to a simulated data set in order to show its efficacy and to ultrafine particle number concentration in Helsinki, Finland, to show its use in real world problems.

\end{abstract}

\begin{keyword}
{autoregressive errors} \sep {semi-parametric regression} \sep {Bayesian inference} \sep {generalised additive model} \sep {ultrafine particles} \sep {aerosols}
\end{keyword}

\end{frontmatter}

\section{Introduction}
Continuously measured time series may exhibit regular temporal trends, non-linear dependence on covariates (and interactions thereof) and autocorrelation. This motivates the desire for flexible regression models which are able to take these features into account without specifying the functional form of the relationship \emph{a priori}.

The use of splines for non- and semi-parametric modelling smooth curves and surfaces \citep{silverman85}, time series \citep{wahba90} and non-linear covariate effects \citep{lin99} is well established \citep*{ruppertwandcarroll09}. Splines have simple to construct bases \citep{deboor78} and can be extended to include smoothness penalties, periodic bases, and interactions \citep{eilersmarx2010}. The flexibility of splines has led to their adoption as bases for Generalised Additive Models \citep{lin99, langbrezger04, wood06}.

\citet*{harveykoopman93} describe a spline based model which flexibly models periodic trends by allowing the spline coefficients to evolve according to a random walk. This model admits the use of covariates but the error terms are assumed independent and identically distributed.

An important consideration when seeking to fit a model for forecasting time series data is that the residuals may display some degree of autocorrelation. Rather than fitting the mean predictor with an assumption of independent and identically distributed residuals and then performing \emph{post hoc} analysis of the residual autocorrelation, including the autocorrelation in the modelling will ensure that the estimates of the model parameters have taken the autocorrelation into account, leading to more realistic forecasting \citep*{Chib93, molgaard2011}.

Traditional time series approaches \citep[e.g.][]{boxjenkins} are able to account for error structures other than independent and identically distributed errors through the specification of the Moving Average component of an ARIMA model \citep{venables}. Seasonality can be modelled in the ARIMA framework by including a term of the form $(1-L^s)$, where $L$ is the lag operator and $s$ is the length of the period in terms of the frequency of the time series. The inclusion of this term accounts for seasonality by effectively removing the seasonal trend at lag $s$. Seasonal decomposition with loess \citep*{stl, loess} can perform interpolation and smooth estimation of seasonal trends with local polynomial regression. A common use of these seasonal decomposition models is to seasonally adjust a time series in order to examine the residual trends \citep*{x12arima}, rather than examining the temporal trends themselves. Forecasting from these seasonally decomposed local regression models is possible, with ARIMA modelling providing the forecasts of the departures from the seasonal patterns.

One of the most flexible and easily implemented bases for the Generalised Additive Model is the penalised B-spline \citep{bspline96}. Despite the development of the GLM with autocorrelated errors \citep{Chib93} and the simplicity of the B-spline there does not appear to be an attempt made to incorporate both of these modelling approaches. A Bayesian nonparametric regression model with autocorrelated errors was proposed and implemented by \citet*{smithwongkohn98} which uses cubic regression splines rather than penalised splines. Penalised B-splines are attractive because when a large number of spline basis functions are used, any excessive wiggliness is penalised and a smoother fit obtained.

In this paper a method which combines spline-based semi-parametric regression modelling with the modelling of auto-regressive errors is proposed. It is possible to define a model for autoregressive (AR) errors in WinBUGS \citep*{winbugs, haypettitt01} but incorporating them with complex spline-based regression fixed effects terms is cumbersome in this modelling environment \citep*{bayesianwinbugs05}. The approach outlined below is a combination of the work of \citet{molgaard2011} and \citet*{Clifford2011}.

Section \ref{sec:meth} features a review of the Generalised Additive Model (GAM) and autocorrelated errors. The use of penalised splines is outlined, including how to construct penalties for multivariate bases and how to ensure the identifiability of the model (as the spline basis has full rank). Gibbs sampling for the spline coefficients and parameters for the autoregressive errors are discussed, as is the Metropolis sampler for the hyperprior which penalises the wiggliness of the splines. Forecasting with AR errors is also discussed, including posterior checks for predictive performance.

In Sections \ref{sec:case} and \ref{sec:results} the model is applied to two data sets. The first is a set of simulated time series data with a two dimensional (2D) covariate and autocorrelated errors added. This shows that the model recovers the temporal trend, autocorrelation of the errors and the 2D covariate effect. The posterior covariance of the sampled residuals are analysed to show the reduction in autocorrelation. The full regression and forecasting model is applied to observations of PNC and meteorology recorded in Helsinki.

In Section \ref{sec:conclusion} the features of the proposed model are reviewed in light of the case studies and suggestions made for extensions to the modelling approach presented.

\section{Methodology}\label{sec:meth}
Bayesian regression modelling can be conceptualised as starting with a set of assumptions about model parameters (prior belief) and using collected data to update those assumptions (to obtain the posterior belief). Mathematically, the prior beliefs about the parameters, \vec{\theta}, such as their mean and variance, are represented as a distribution, $p\left(\vec{\theta} \right)$. The regression model and data are represented by a distribution where the likelihood of the data ($\mat{X}$) is conditioned on the parameters, $p\left( \mat{X} \pipe \vec{\theta} \right)$. The posterior is obtained, through Bayes' rule by multiplying the prior and the likelihood, $p\left( \vec{\theta} \pipe \mat{X} \right)$, and represents a probability distribution for the parameters, conditioned on the observed data \footnote{For a more thorough review of Bayesian statistics, see \citet{gelman}}.

\citet{Chib93} specifies a Bayesian linear model in which the responses have either a Gaussian or $t$ likelihood and the residuals are assumed to be autocorrelated rather than independent and identically distributed. In this section, an additive model with autocorrelated errors is developed by using splines to estimate non-linear covariate effects \citep{hastietib}. The benefits of spline regression over polynomial regression have been discussed by \citet{ruppertwandcarroll} and \citet{bayesianwinbugs05}.

Current software packages for Bayesian semi- and non-parametric Bayesian regression, such as R-INLA \citep{inlapackage}, mgcv\footnote{while not strictly Bayesian, mgcv gives Bayesian summaries} \citep{mgcv} and BayesX \citep{bayesX}, do not deal with autocorrelated residuals.

The methodology is implemented in MATLAB. The regression model is fit to the data which is observed and whose lagged data has also been observed. That is, we omit any data where any of $y_i$ or its lagged values are missing (represented in MATLAB as \texttt{NaN}, ``not a number'').

\subsection{Generalised Additive Models}\label{sec:gam}
Generalised Additive Models (GAMs) can be treated as equivalent to semiparametric regression with the Generalised Linear Mixed Model (GLMM) where the nonparametric smoothers, such as splines, are the random effects \citep*{ruppertwandcarroll,inlaglmm07}. Alternatively, the basis matrix for the parametric part of the model can be augmented with the basis matrices for said smoothers, providing for the conception of the GAM as replacing the linear terms in the GLM with additive smoothers \citep{hastietib}.

A GAM can therefore be expressed as
\begin{align*}
\mu_i & = \beta_0 + f_1 \left(x_{i1} \right) + f_2 \left(x_{2i} \right) + \ldots + f_M \left(x_{Mi} \right) \\
\vec{\mu} & = \mat{X}\vec{\beta}
\end{align*}
for $M$ covariates, $x_{m\ast}$, each with their own univariate spline, $f_m(x_{m\ast})$, and a link function, $g(\cdot)$, as in the GLM. The design matrix of the GLM, $\mat{X}$, now represents the spline design matrix. It has as many rows as there are observations and $\sum_{m=1}^M d_m$ columns, where $d_m$ is the number of basis splines used in the construction of $f_m(\cdot)$.

The GAM can thus be written in a form which is equivalent to the formulation of a GLM, but with splines replacing the linear terms in the predictor,
\begin{equation}
g \left(  \vec{y} \right) = {\mat{X}} {\vec{\beta}} + \vec{\varepsilon} \label{eq:glm}.
\end{equation}

\subsection{Penalised splines}
\subsubsection{B-splines}\label{sec:bsplines}
The B-spline basis is built according to a recursive algorithm \citep{deboor77,bspline96}. A group of step functions, each of which is non-zero between two adjacent knots, $t_{j-1}$ and $t_j$, are operated on to build a group of continuous linear functions which are non-zero between three adjacent knots such that the basis functions overlap. This process is repeated until the desired basis order is obtained. 

The order $k$ B-spline for covariate $x$ 
\begin{multline}
\frac{B_{j,k}(x)}{t_{j+k}-t_{j}} = \frac{x - t_j}{t_{j+k-1}-t_{j}} \frac{B_{j,k-1}(x)}{t_{j+k-1}-t_{j}} + \\  \frac{t_{j+k} - x}{t_{j+k}-t_{j}} \frac{B_{j+1,k-1}(x)}{t_{j+k}-t_{j+1}} \label{eq:bs}
\end{multline}
on a grid of knots $t$ with $B_{j,0}(x) = 1$ between knots $t_{j-1}$ and $t_j$. That is, the zero-order B-spline basis is constant between successive knots and higher order bases are created according to the recurrence relation.

A cubic B-spline basis function consists of three quadratic polynomials which are piecewise continuous and smooth in the intervals between four neighbouring knots and the basis outside these knots is identically zero. Each B-spline basis function has compact support (i.e. are non-zero only on an open interval) and so allow the flexible modelling of non-linear effects by responding to local changes in the relationship between the corresponding covariate and the response.

The basis splines are calculated once for each non-linear function to be approximated and so the choice of the order of spline doesn't affect the complexity or speed of the calculations in the MCMC simulation, although the number of knots chosen will do so. The choice of a particular order of spline basis can therefore be made according to the desired continuity, smoothness and differentiability properties of the fitted smooth. A B-spline basis function will have non-zero derivatives up to and including the order of the basis spline, e.g. a first order spline will be once differentiable.

\begin{figure}[ht]
  \centering
\includegraphics[width=\linewidth]{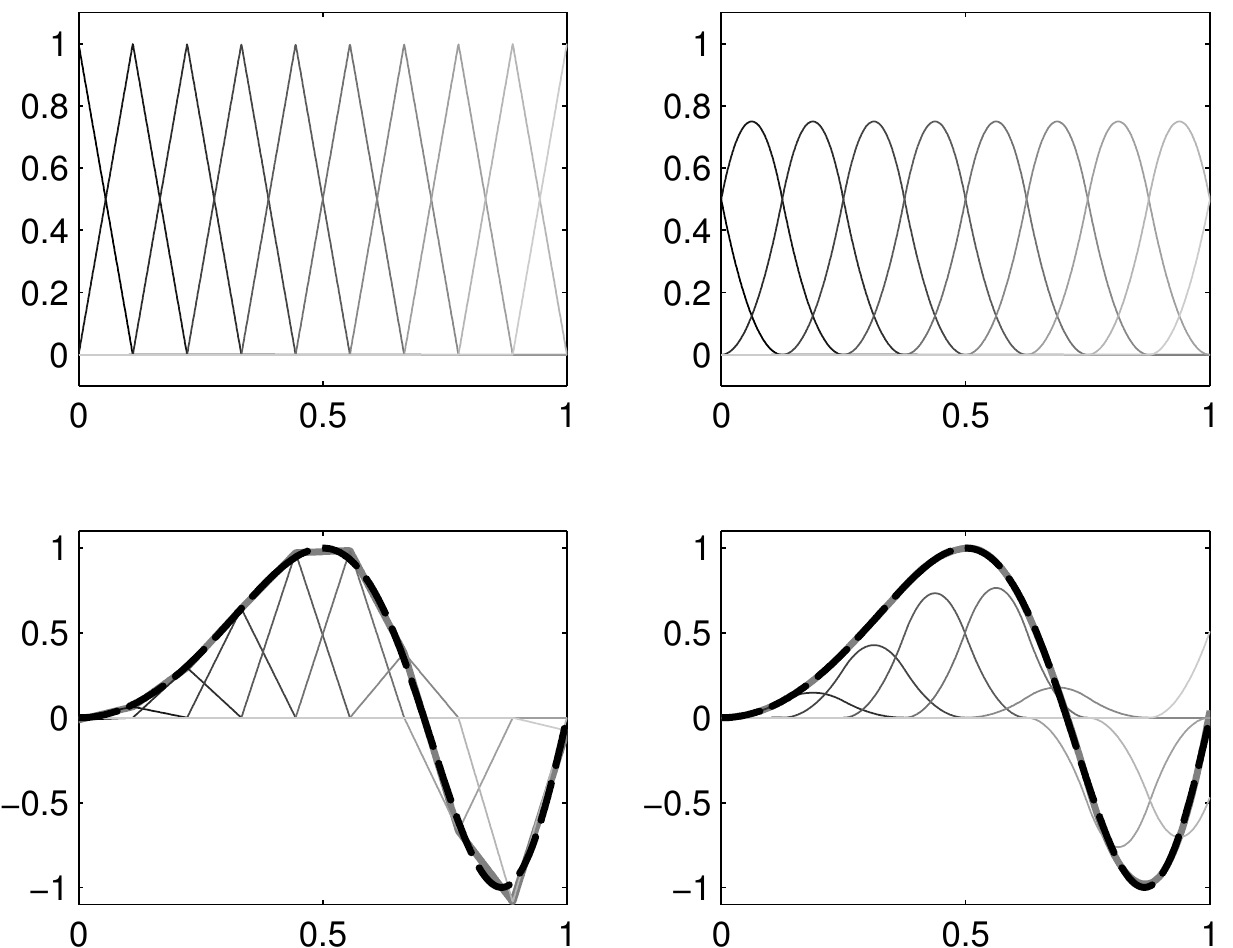}
\caption{Example B-splines. Top row: ten first order B-splines, ten second order B-splines. Bottom row: Use of B-splines to approximate $y = \sin \left( 2 \pi x^2 \right)$ where the true function is a dashed black line and the approximation is a solid grey line.}    
\label{fig:bsplines}            
\end{figure}

Figure \ref{fig:bsplines} shows ten first order B-splines, ten second order B-splines and a linear combination of the same to approximate the function $y = \sin \left( 2 \pi x^2 \right)$. The first order splines are piecewise continuous and that the second order splines are piecewise smooth. Each individual basis function is coloured a different shade of grey; the colouring scheme is consistent across each of the four figures. The approximation is shown as a thick grey line and the true function is shown as a thick, black, dashed line.

\begin{figure}[ht]
\centering
\includegraphics[width=\linewidth]{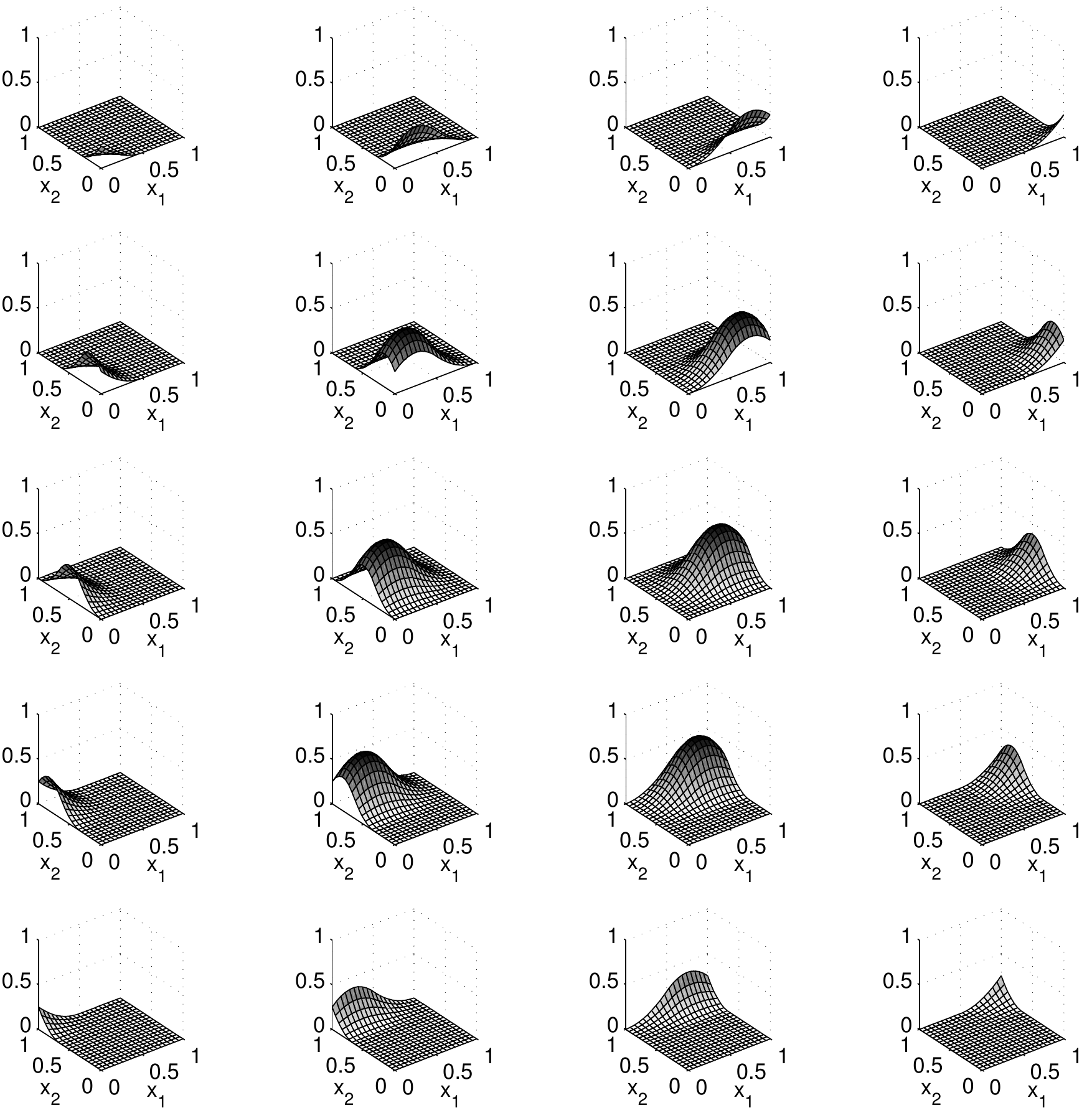}
\caption{Tensor product of two B-splines. $x_1$ is modelled with four second order B-splines, $x_2$ with five second order B-splines; the tensor product thus has 20 basis elements. Shading is according to the value of the joint basis. Each image represents a 2D basis spline in the joint basis as a matrix, the spline basis matrix will consist of these matrices reshaped as column vectors.} 
\label{fig:bs2d}
\end{figure}

Figure \ref{fig:bs2d} shows a 2D tensor basis of B-splines.

A cyclic B-spline basis can be defined by adding the basis splines with non-zero value at the edge of the univariate covariate space such that the basis is piecewise continuous at the boundary and any order preserving permutation of the knots on the periodic covariate space results in a valid B-spline basis.

\subsubsection{Penalisation}
A smoothing penalty can be imposed to ensure that the fitted B-splines are not too ``wiggly''. \citet*{bspline96} introduce the penalised B-spline, or ``P-spline'', in a frequentist context, modifying the normal equations to include a penalty matrix based on discrete differences between the spline coefficients. 

Let $\mat{D} = \mat{D}(k)$ be a discretised operator equivalent to the $k$th derivative of the identity matrix with dimension equal to the number of coefficients in the B-spline basis of interest. This matrix will be rank deficient, with the deficiency in rank being equal to the order of the derivative.  Applying the differential operator to a vector $\vec{\beta} \in \mathbb{R}^d$, where $d$ is the number of basis elements in a univariate spline whose coefficients are $\vec{\beta}$, yields a vector with a value of the discretised value of that differential operator evaluated at $\vec{\beta}$, and has a dimension equal to the number of rows of the differential operator, i.e. for $\vec{\beta} \in \mathbb{R}^d, \mat{D} \in \mathbb{R}^{ (d-k) \times d}$, so that $\mat{D} \vec{\beta} \in \mathbb{R}^{ (d-k) \times 1}$.

As $\mat{D} \vec{\beta}$ is the application of the discretised differential operator, $\vec{1}^T \mat{D} \vec{\beta} = \sum_{i=1}^{d-k} \mat{D}_{i,*} \vec{\beta}$, is the sum of the discretised derivatives. The continuous analogy is $\int f^{(k)} (x) \, \mathrm{d}x$. Then
\[
\left( \vec{\beta}^T \mat{D}^T\right) \left(\mat{D} \vec{\beta}\right) = \int \left[ f^{(k)} \right]^2 \, \mathrm{d}x
\]
and is the discretised sum of the squares of the derivatives \citep{wood03}. The quantity to be minimised in the penalised B-spline, then, is the sum of the square of the derivative \citep{bspline96}.

\citet{langbrezger04} reformulate this penalty matrix as a zero mean conjugate multivariate normal prior on the B-spline coefficients, $\vec{\beta} \in \mathbb{R}^{d \times 1}$,
\begin{align}
\left( \vec{\beta} \pipe \lambda \right) & \propto \lambda^{(d-k)/2} \exp \left( -\frac{\lambda}{2} \vec{\beta}^T \mat{K} \vec{\beta}  \right) \label{eq:pspline} \\
\lambda & \sim  \Gamma(1,b) \label{eq:gammaprior}
\end{align}
where $\lambda$ is the smoothing parameter which penalises the ``wiggliness'' of the resulting spline and $\mat{K}  \in \mathbb{R}^{d \times d}$ is the matrix form of the discretised wiggliness penalty as in \citet{bspline96}. The parameter for the scale of the prior on $\lambda$, $b$, is chosen such that the prior has a small variance and is concentrated around $\lambda = 0$. The prior for $\vec{\beta}$ is improper, but a non-diffuse prior on $\lambda$ ensures that the posterior for $\vec{\beta}$ is proper. The rank deficiency of the penalty matrix is overcome by adding a small amount to the diagonal of $\mat{K}$ of no more than about 0.001 (the diagonal values of $\mat{K}$ are at least 2, the first order penalty case).

There are other ways to penalise the B-spline coefficients than the P-spline approach of \citet{eilersmarx2003}. These include the approach of \citet{osullivan88} and its Bayesian analogue, developed by \citet{wand08}.

\subsubsection{Interaction terms}
\citet*{harveykoopman93} describe a spline based model which is able to flexibly model periodic trends by allowing the spline coefficients to evolve according to a random walk. This model is applied to a joint model of daily and hourly trends and treats the coefficients of the daily trend spline as evolving according to a random walk model from day to day. While this is an obvious choice for modelling an interaction between a factor and a continuous covariate, it does not make sense for two continuous covariates.

A tensor product of the bases for two univariate B-spline matrices can be constructed in order to fit an interaction term of those covariates \citep*{eilersmarx2003,marxeilers2005}. Penalisation of the coefficients for this tensor basis is achieved by forming the Kronecker tensor product, denoted $\otimes$, of the identity matrix with size equal to the number of basis elements and the univariate penalty matrix for each univariate spline. This leads to a formulation of the smoothing penalty prior in terms of two penalty matrices, one along each axis of the covariate space corresponding to each of the two covariates. The prior takes the form
\begin{equation}
\begin{aligned}
\left( \vec{\beta} \pipe \lambda_1, \lambda_2 \right) & \propto \exp \left( -\frac{\lambda_1}{2} \vec{\beta}^T \mat{P}_1 \vec{\beta} -\frac{\lambda_2}{2} \vec{\beta}^T \mat{P}_2 \vec{\beta}  \right) \label{eq:betaprior} \\
\lambda_1, \lambda_2 & \sim \Gamma(1,b) \\
\mat{P}_1 & =  K_1 \otimes \mat{I}_{d_2} \\
\mat{P}_2 & =  \mat{I}_{d_1} \otimes K_2  
\end{aligned}
\end{equation}
for $d_1$, $d_2$ the number of basis elements for covariates 1 and 2 respectively. The matrices $\mat{P}_1$ and $\mat{P}_2$ are square matrices of dimension $d_1 \times d_2$, the first of which is block diagonal, with each block composed of tridiagonal matrices, corresponding to the penalties on the first covariate's basis (indexed sequentially). The second, $\mat{P}_2$, has a banded structure, corresponding to the penalties on the second covariate's basis (where the indexing is sequential, $d_1$ at a time). An example of a 2D tensor product of B-splines is shown in Figure \ref{fig:bs2d} in the appendix.

The above tensor product formulation with a penalty for each univariate basis reflects the possibility that a different amount of smoothing may apply in each direction, i.e. that smoothing is not isotropic. While R-INLA provides a \texttt{rw2d} GMRF latent model for two-dimensional smooth terms, the single precision parameter of the GMRF (equivalent in formulation to a univariate P-spline prior) does not reflect that there may be a different amount of smoothing required in each direction. This multiple penalty is addressed in BayesX, along with an interaction between the smoothing terms \citep{bayesX}.

The random walk time evolving spline of \citet{harveykoopman93} can be recovered by using a first order random walk P-spline penalty matrix as the precision matrix for a factor term.

\subsection{Parametric terms}
In this section, the term ``parametric'' denotes regression basis functions with a parametric form specified \emph{a priori}. These parametric terms, such as linear terms, polynomials and sinusoidal functions, will be modelled as fixed effects with Gaussian priors (multivariate where appropriate), according to the method outlined by \citet{Chib93}.

In section \ref{sec:bsplines} the periodic B-spline basis was briefly discussed. Another choice of basis for semi-parametric modelling of periodic functions is the Fourier series basis, consisting of $\sin\left( \frac{c \pi x}{L} \right)$, $\cos\left( \frac{c \pi x}{L} \right)$, $c = 1\ldots C$. The full Fourier series basis also includes an intercept term which should be omitted from the regression model to ensure identifiability of the overall mean. In the case of the splines section \ref{sec:identifiability} discusses how identifiability is enforced by centring the splines; the $\sin$ and $\cos$ terms are already centred.

The sinusoidal functions may be treated as any other parametric (e.g. linear) term in the regression equation. It is appropriate to assume that the Fourier coefficients are uncorrelated since by definition the Fourier basis is orthogonal. An appropriate prior, therefore, is a weakly informative multivariate normal with zero mean and a covariance matrix whose structure is diagonal.

\citet{heilerfeng00} provide an example of using the Fourier series basis to approximate temporal trends. As with all variable selection problems, the question is how many terms must be included? A difficulty of using Fourier series is that using too few terms constrains the ``wiggliness'' of the resulting approximated function whereas using terms with too high frequency may lead to spurious oscillations unless the prior shrinks the coefficients of higher frequency oscillations to zero as in \citet{lenk99}. Instead of this approach, the use of periodic B-splines with a smoothing penalty is adopted. Here it is merely pointed out that the Fourier basis may be used, as was done by \citet{molgaard2011}.

\subsection{Factor terms}
Factor terms may arise as random effects in a GAMM, e.g. a site-specific mean in multi-site data, and may be treated by coding all levels of the factors to an integer, as in R-INLA. For a factor basis matrix with sequential integers indexing the factors, $\mat{F} \in \mathbb{R}^{n \times J}$, for covariate $\vec{x} \in \mathbb{R}^n$, the element $F_{i,j} = 1$ if $x_i = j$ ($j \in 1, 2, \ldots, J$).

For the coprecision of the factor term, $\mat{Q}_F$, it is simple to assume that the factors are independent and identically distributed, so that $\mat{Q}_F = \lambda_F \mat{I}$ for precision parameter $\lambda_F$. If there is some known structure in the factors or if a factor is being used to describe a zero order spline (constant between knots with knots spaced between factor indices) then a smoothing penalty prior may be used, as for a P-spline.

The factor terms are centred around zero with the same identifiability constraint as in \ref{sec:identifiability}.

\subsection{Autocorrelated residuals}\label{sec:ar}
Many time series are highly autocorrelated. Any temporal variation not explained by a periodic spline basis (e.g. hour of day, day of year) in the proposed model can treated by modelling the time series of errors, ${\varepsilon}_i$, with an autoregression model,
\begin{align}
\phi (L) \varepsilon_i & = u_i  \label{eq:ar} \\
u_i & \sim \mathcal{N} \left(0, \sigma^2 \right). \nonumber
\end{align}
where $L$ is the lag operator and $\phi (L)$ is a $p$-degree polynomial in the lag operator
\begin{equation*}
\phi (L) x_i  = x_i + \phi_1 x_{i-1} + \phi_2 x_{i-2} + \ldots + \phi_p x_{i-p}
\end{equation*}
describing the autocorrelation structure of the residuals in \ref{eq:glm} \citep{boxjenkins}.

\citet{Chib93} suggests the use of a multivariate normal prior for the coefficients of the above polynomial. A censoring condition, that the absolute sum of the coefficients is strictly less than one, is applied in order to ensure that the error process is stationary. The prior is
\[
\vec{\phi} \propto \, \mathcal{MVN} \left( \vec{\phi}_0 , \mat{\Phi}_0^{-1} \right) \; : \; {\left\vert\vec{1}^T \vec{\phi}\right\vert < 1}.
\]
where the prior mean, $\vec{\phi}_0$, and precision, $\mat{\Phi}_0 \in \mathbb{R}^{p \times p}$, are set such that the prior for the autoregression parameters is weakly informative. The prior for the variance of the autoregressive errors is a conjugate inverse Gamma,
\[
\sigma^2 \sim \, \mathcal{IG} \left( \frac{v_0}{2}, \frac{\delta_0}{2} \right).
\]

By using an autoregressive error structure, forecasting from the model will give a more realistic representation of the relationship between successive observations/forecasts. If the residuals are independent, this will manifest as the zero vector being contained within the multivariate credible set of $\vec{\phi}$.

By sampling the residuals in the MCMC scheme and using these for forecasting, a distribution for each forecast is readily obtainable. This will be especially important when making a forecast which is dependent on a previously forecast observation. Ignoring the uncertainty in the distribution of the forecast value $\widehat{y}_{t+1}$ would produce an estimate of $\widehat{y}_{t+2}$ which assumed that $\hat{y}_{t+1}$ was observed rather than forecast.

\subsection{Estimation via MCMC}
The form of the GAM with log link, Gaussian likelihood and autocorrelated residuals is
\begin{equation}
\begin{aligned}
 \log y_i = & \, \mu_i + \varepsilon_i \\
 \mu_i = & \, \mat{X}_{\ast,i}\vec{\beta} \\
\phi (L) \varepsilon_i = & \, u_i \\
u_i \sim & \, \mathcal{N}\left(0, \sigma^2 \right)
\end{aligned}\label{eq:gamar}
\end{equation}
where $\mat{X}_{i,\ast}$ is row $i$ of matrix $\mat{X}$.

A model for a Gaussian hierarchical linear model with autoregressive errors is given by  \citet{molgaard2011}, following the formulation of \citet{Chib93}. The priors for $\sigma^2$, $\vec{\beta}$ and $\vec{\phi}$, in this formulation are
\begin{equation}
\begin{alignedat}{2}
\left( \vec{\beta} \pipe \sigma^2 \right) \sim & \, \mathcal{MVN} \left( \overline{\vec{\beta}}_0, \sigma^2 \mat{A}_0^{-1} \right) \\
\sigma^2 \sim & \, \mathcal{IG} \left( \frac{v_0}{2}, \frac{\delta_0}{2} \right) \\
\vec{\phi} \propto & \, \mathcal{MVN} \left( \vec{\phi}_0 , \mat{\Phi}_0^{-1} \right) \; : \; {\left\vert\vec{1}^T \vec{\phi}\right\vert < 1}.
\end{alignedat} \label{eq:chibpriors}
\end{equation}
The terms $\phi_0$ and $\overline{\vec{\beta}}_0$ are the respective mean values for the weakly informative priors for $\vec{\phi}$ and $\vec{\beta}$, taken to be \vec{0} in each case. The values for $v_0$ and $\delta_0$ are taken to be $-k$ and $0$, respectively, as in \citet{Chib93}. The prior precision matrix for the autoregressive error model, $\mat{\Phi}_0$, is set as $10^{-6}\mat{I}_p$ so that the prior is weakly informative, such. The prior precision matrix for $\vec{\beta}$, $\mat{A}_0$ is discussed below.

\citeauthor{Chib93} sets the prior precision of $\vec{\beta}$, $\mat{A}_0$, as a scalar multiple of an identity matrix such that the prior is weakly informative. Here, however, $\mat{A}_0$ is a block diagonal with each block containing a smoothness penalty precision matrix, $\lambda_j \mat{K}_j$, for a spline term. The $\lambda$ are therefore hyperparameters for the prior on $\vec{\beta}$. Any parametric terms to be included in the model can be treated similarly, by including their prior precision as a series of individual elements on the diagonal (in the case that they are independent), a block consisting of a diagonal matrix (for sets of independent, identically distributed parameters) or a block structured as a symmetric positive definite matrix (if there is some known correlation structure). The prior for $\vec{\beta}$ is now is very informative due to the spline penalties contained within $\mat{A}_0$.

The posterior distributions for the linear predictor terms, $\vec{\beta}$, autoregressive coefficients, $\vec{\phi}$, and model variance, $\sigma^2$ are given in Table \ref{tab:posteriors}.

\begin{table*}[htb]
\begin{align}
\left( \vec{\beta} \pipe \vec{y}, \sigma^2, \vec{\phi} \right) \sim & \, \mathcal{MVN}\left( \mat{\Lambda}_{\beta} \left( \mat{A_0} \overline{\vec{\beta}}_0 + {\mat{X}^{\ast}}^T \mat{y}^{\ast} \right) , \sigma^2 \mat{\Lambda}_{\beta} \right) \label{eq:beta}\\
\left( \sigma^2 \pipe \vec{y}, \vec{\beta}, \vec{\phi} \right) \sim & \, \Gamma^{-1} \left( \frac{n - p + v_0 + k}{2}, \frac{\delta_0 + \mat{Q}_{\beta} + \left\| \mat{y}^{\ast} - \mat{X}^{\ast}\vec{\beta} \right\|_2^2 }{2} \right) \label{eq:sigma} \\
\left( \vec{\phi} \pipe \vec{y}, \vec{\beta}, \sigma^2 \right) \sim & \, \mathcal{MVN}  \left( \mat{\Lambda}_{\phi}  \left( \mat{\Phi}_0 \phi_0 + \sigma^{-2} \mat{E}^T \vec{\varepsilon} \right) , \mat{\Lambda}_{\phi} \right) \label{eq:phi}
\end{align}
where
\begin{align*}
\mat{\Lambda}_{\beta} = & \, \left(\mat{A}_0 + {\mat{X}^{\ast}}^T \mat{X}^{\ast} \right)^{-1} \\
\mat{Q}_{\beta} = & \, \left( \vec{\beta} - \overline{\vec{\beta}}_0 \right)^T \mat{A_0} \left( \vec{\beta} - \overline{\vec{\beta}}_0 \right) \\
\mat{\Lambda}_{\phi} = & \, \left(\mat{\Phi}_0 + \sigma^{-2} \mat{E}^T \mat{E} \right)^{-1}.
\end{align*}
\caption{Posterior densities for GAM with AR residuals}
\label{tab:posteriors}
\end{table*}

The matrix $\mat{X}^{\ast}$, and the vector $\vec{y}^{\ast}$ are formed by applying the general lag polynomial $1 - \phi(L)$ to $\mat{X}$ and $\vec{y}$ respectively, reducing the number of rows by $p$, the number of lags included in the autoregressive residual term. That is,
\begin{multline}
 \mat{X}^{\ast} = \mat{X}_{1 \ldots n-p,\ast} - \phi_1 \mat{X}_{2 \ldots n-p+1,\ast} - \\ \phi_2 \mat{X}_{3 \ldots n-p+2,\ast} \ldots - \phi_p \mat{X}_{p+1 \ldots n,\ast}
\end{multline}
and similarly for $\vec{y}^{\ast}$.

$\mat{E} \in \mathbb{R}^{(n-p) \times p}$ contains, as its columns, the residuals, $\vec{\varepsilon}$, at each of the $p$ lags, such that the $i^{\textnormal{th}}$ row is $\left[ \varepsilon_{i-1}, \ldots, \varepsilon_{i-p} \right]$,
\begin{equation}
\mat{E} =  \left[ \vec{\varepsilon} \pipe  L \vec{\varepsilon} \pipe  L^2 \vec{\varepsilon} \pipe  \cdots \pipe  L^p \vec{\varepsilon} \right].
\end{equation}
In \ref{eq:sigma}, $\left\| \cdot \right\|_2$ represents the Euclidean 2-norm.

Rather than first fitting the linear predictor with the assumption of independent, identically distributed errors and then fitting an autoregressive model for the residuals so that $\vec{\phi}$ is conditioned on $\vec{\beta}$ but not vice versa, this comprehensive model allows simultaneous estimates of $\vec{\beta}$ and $\vec{\phi}$.

\begin{figure*}[ht]
\centering

\tikzset{blanknode/.style={draw=none,fill=none,circle,minimum size=3.5em}}

\begin{tikzpicture}[scale=1.25]
 \tikzstyle{every node}=[fill=white,draw=black,thick,circle,minimum size=2.75em] 
\tikzstyle{every edge}=[black,-latex,thick,draw]
\node (y) at (0,-0.5) {$y_i$} ;
\node (mu) at (1,0) {$\mu_i$}  edge (y);
\node (X) at (2,0.5) {$\mat{X}$}  edge (mu);
\node (beta) at (2,-0.5) {$\vec{\beta}$} edge (mu);
\node (beta0) at (3,-1) {$\overline{\vec{\beta}_0}$} edge (beta) ;
\node (A0) at (3,0) {$\mat{A}_0$} edge (beta) ;
\node (K) at (4,0.5) {$\mat{K}_j$} edge (A0);
\node (lambda) at (4,-0.5) {$\lambda_j$} edge (A0);
\node (b) at (5,0) {$b$} edge (lambda);
\node (eps) at (-1,0) {$\varepsilon_i$}  edge (y);
\node (phi) at (-2,0.5) {$\vec{\phi}$} edge (eps);
\node (u) at (-2,-0.5) {$u_i$} edge (eps) ;
\node (phi0) at (-3,0) {$\vec{\phi}_0$} edge (phi) ;
\node (capphi0) at (-3,1) {$\mat{\Phi}_0$} edge (phi) ;
\node (sig) at (-3,-1) {$\sigma^2$} edge (u) ;
\node (v0) at (-4,-1.5) {$v_0$} edge (sig) ;
\node (delta0) at (-4,-0.5) {$\delta_0$} edge (sig) ;
\node[blanknode] (Kdummy) at (K) {};
\node[blanknode] (lambdummy) at (lambda) {};
\begin{pgfonlayer}{background}
\filldraw [fill=black!5,draw=black,thick,rounded corners=1mm,inner sep=0pt] (Kdummy.north -| lambdummy.west) rectangle (lambdummy.south -| Kdummy.east);

   \end{pgfonlayer}
\node[fill=none,draw=none,above=of Kdummy,yshift=-1.5cm] {$j$=1..nsplines};
\end{tikzpicture}

\caption{Directed acyclic graph of the model in \ref{eq:gamar} with priors \ref{eq:chibpriors}, \ref{eq:pspline}, \ref{eq:gammaprior} and \ref{eq:betaprior}.}
\label{fig:dag}
\end{figure*}
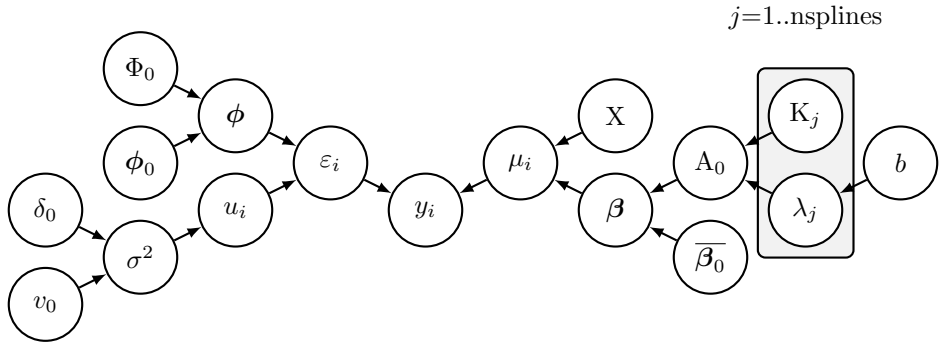

The directed acyclic graph of the model is given in Figure \ref{fig:dag}. Sampling is performed by successively sampling from the marginals of $\vec{\beta}^{(t)}$ from \ref{eq:beta} (then correcting $\vec{\beta}^{(t)}$ according to \ref{eq:betacorrect1} and \ref{eq:betacorrect2}),  $\vec{\phi}^{(t)}$ from \ref{eq:phi}, ${\sigma^2}^{(t)}$ from \ref{eq:sigma} and  sample $\vec{\lambda}^{(t)}$ from \ref{eq:lambda} and \ref{eq:lambda2d} with a Metropolis-Hastings step.

\subsubsection{Identifiability of spline coefficients}\label{sec:identifiability}
A $d$-dimensional spline basis $\mat{B}$ will have full rank and therefore the matrix $\left[ \vec{1} | \mat{B}\right] \in \mathbb{R}^{n \times (d+1)}$ is rank deficient. In order to enforce identifiability the B-spline parameters are centred at each step of the Gibbs sampler such that the sum of the marginal effect of that spline is zero \citep{wood03} and centering constant is transferred to the intercept term in the model, $\beta_0$.

At step $t$ in the Gibbs sampler, for each block of the covariate matrix $\mat{X} \in \mathbb{R}^{N \times d}$ corresponding to a B-spline, indexed within $\mat{X}$ by $\textnormal{ind}$ we set
\begin{equation}
\begin{aligned}
\vec{\beta}_{\textnormal{ind}}^{(t)} \leftarrow & \vec{\beta}_{\textnormal{ind}}^{(t)} - \delta_{\beta} \vec{1} \\
\beta_{0}^{(t)} \leftarrow & \beta_{0}^{(t)} + \delta_{\beta}.
\end{aligned}\label{eq:betacorrect1}
\end{equation}
where the centring constant, $\delta_{\beta}$, is calculated as that which satisfies
\begin{align}
0 = & \sum\limits_{i=1}^N \left( \mat{X}_{i,\textnormal{ind}} \vec{\beta}_{\textnormal{ind}}^{(t)} - \mat{X}_{i,\textnormal{ind}} \vec{1} \delta_{\beta} \right)\nonumber\\
\delta_{\beta} = & \frac{\sum\limits \mat{X}_{\ast,\textnormal{ind}} \vec{\beta}_{\textnormal{ind}}^{(t)} }{\sum\limits_{i=1}^N \sum\limits_{j \in \textnormal{ind}} \mat{X}_{i,j}}.\label{eq:betacorrect2}
\end{align}

\subsubsection{Lack of support}\label{sec:support}
It is possible that B-spline knots will be placed in a region of the covariate space where there are no observed values, especially when dealing with interaction terms. For a basis spline defined in a region of no support (i.e. there are no observations in the relevant region of the covariate space) the posterior of the corresponding coefficient is undefined. In these cases, a sample will be taken from the prior. Due to the informative nature of the penalty prior, the posterior variance of this parameter will be much smaller than if the spline was unpenalised. The corresponding marginal effect of the entire spline will have a wider credible interval in the region of this spline's support.

An alternative is to use a spline basis with global support for the covariate responsible for the lack of support, such as the semi-parametric low rank thin plate spline described by \citet{ruppertwandcarroll} and \citet{bayesianwinbugs05} as
\[
f(x_i) = \sum\limits_{j=1}^J \beta_j x_i^j + \sum\limits_{k=1}^{K} \gamma_k \left\vert x_i - \kappa_k \right\vert^m
\]
where $J < m$ is the order of the parametric fixed effect polynomial, $m$ is the order of the random effects polynomial splines and the $\kappa_k$ are knots placed at the $k/(K+1)$ quantiles of $x$. A suitable prior for $\left( \beta_1, \ldots, \beta_J, \gamma_1, \ldots \gamma_K \right)$ is a weakly informative multivariate normal with an identity precision matrix.

\subsubsection{Sampling spline penalties}
In order to sample the P-spline penalty correctly we need to extend the Gibbs sampler to also sample the $\lambda$, which are hyperparameters for the prior distribution for $\vec{\beta}$. The penalty matrices $\mat{K}$ and $\mat{P}$, suggested by \citet{langbrezger04}, however, have a determinant of zero and thus the posterior of $\lambda$ is undefined. We avoid this by adding a small value to the diagonals of these matrices to ensure that they are non-singular and thus invertible, e.g. the penalty becomes $\lambda \left( \mat{K} + 10^{-5} \mat{I}\right)$.

For the one-dimensional splines it is rather simple to find the conditional distribution of the $\lambda$ hyperparameters when $\mat{K} \in \mathbb{R}^{d \times d}$ has non-zero determinant. The posterior is derived as
\begin{align}
p\left(\lambda \pipe \vec{\beta}\right) & \propto p\left(\vec{\beta} \pipe \lambda\right)p(\lambda) \\
& = \mathcal{MVN} \left(\vec{\beta} \pipe \vec{0},\left( \lambda \mat{K} \right)^{-1} \right) \Gamma\left( \lambda \pipe 1,b \right) \nonumber  \\
& =  \sqrt{ \frac{\det \left( \lambda \mat{K} \right)}{\left( 2\pi \right)^d} } \exp\left( -\frac{\lambda}{2} \vec{\beta}^T \mat{K} \vec{\beta} \right) be^{-b\lambda} \nonumber \\
& \propto \lambda^{d/2} \exp \left( -\lambda \left(b+\frac{1}{2}\vec{\beta}^T \mat{K} \vec{\beta}\right) \right) \nonumber \\
& \propto \Gamma\left(\lambda \pipe \frac{d}{2} + 1 , b + \frac{1}{2} \vec{\beta}^T \mat{K} \vec{\beta} \right) \label{eq:lambda}.
\end{align}
That is, the Gamma prior, \ref{eq:gammaprior}, is updated by the data in such a way that some smoothing is enforced by the dimension of the spline but the scale parameter, which controls the amount of smoothing (or how likely the smoothing is), is shifted according to the second difference of the parameters. It is still possible to have zero smoothing, but the posterior density is small, so that a lack of smoothing must be justified by the data. 

For two dimensional splines the sampling of $\lambda$ would be as simple as the univariate case if the smoothing was set \emph{a priori} to be the same for each covariate in the tensor product. It is more valid, though, to assume that this is not the case and that there may be more smoothness in one direction than the other, implying two different $\lambda$ parameters. The conditional distribution of these parameters, assuming that the same hyperpriors for the Gamma prior are used for each spline, is
\begin{multline}
p\left(\lambda_1 ,\lambda_2\pipe \vec{\beta}\right) \propto \sqrt{\det \left( \lambda_1 \mat{P_1} + \lambda_2 \mat{P_2} \right) } \left( \lambda_1 \lambda_2 \right)^{a-1} \times \\ e^{ -\frac{1}{2} \vec{\beta}^T \left( \lambda_1 \mat{P_1}+\lambda_2 \mat{P_2} \right) \vec{\beta} } e^{-b \left( \lambda_1 + \lambda_2 \right)} \label{eq:lambda2d}.
\end{multline}

Because the conditional distributions of $\lambda_1$ and $\lambda_2$ are not easy to obtain, we sample them with a Metropolis-Hastings step \citep{hastings70} inside the Gibbs sampler.

\subsubsection{Credible intervals}
Credible intervals of parameter estimates can be calculated quickly from the MCMC samples. For any parametric terms, 95\% credible intervals will be reported. For the coefficients of splines, the credible interval for the entire marginal effect is of interest and so individual credible intervals for each parameter are not so directly interpretable for the spline terms.

\subsection{Forecasting}\label{sec:fore}
The model forecasts in a sequential manner, conditioning each forecast on the observed values and forecast values preceding it according to the autoregressive structure. The uncertainty in forecast values will accumulate, as discussed in Section \ref{sec:ar}.

In Section \ref{sec:csfinland}, to take advantage of the growing set of training data the model is refitted every 20th day. For each day, forecasts are produced one time hour at a time for the coming forty-eight hours, midnight to midnight, using the most recent fit of the training data and the residuals, $\vec{\varepsilon}$, for the past week. These residuals are assumed to be available until noon. The forecast values for each observation made 24 and 48 hours in advance will be stored.

The modelled value of each forecast is obtained by calculating $\widehat{\vec{\mu}} = \mat{X} \widehat{\vec{\beta}}$ for the values to be forecast, based on the observed covariates (in practice, these may be known and the forecasting treated as imputation, they may be forecasts themselves from a model for the covariates, may be generated from a prior, etc.); calculating the residuals, $\widehat{\vec{\varepsilon}}$, for the week before the forecasting is to start; and constructing autocorrelated forecasting errors from the estimates of $\vec{\phi}$. These are all constructed from MCMC output, so distributions of these parameters are all available. The forecasts, $\widehat{y}_i$, are calculated from the MCMC iterations so that
\[
\widehat{y}_i^{(t)} = \, \widehat{\mu}_i^{(t)} + \widehat{{\varepsilon}}_i^{(t)}.
\]
Because both $\widehat{\mu}_i$ and $\widehat{{\varepsilon}}_i$ are normally distributed, $\widehat{y}_i$ is also normally distributed. The posterior density of the forecast values, then, is the empirical density of the samples and is asymptotically normal as the number of samples increases.

For further information, see Sections 2.1.3 to 2.1.5 of \citet{molgaard2011}.

\subsection{Posterior checks}
Convergence of the MCMC chains can be checked by examining the trace of the Markov chains and that the posterior kernel density estimate is unimodal and that its shape is appropriate to the distribution from which it was drawn.

As the posterior for a fitted smooth is a multivariate normal it is important to remember that even though the 95\% CI for a B-spline coefficient (which is marginally normally distributed) may contain zero, it is the joint effect that we are interested in. Indeed, because the spline is centred around zero, it is not unexpected that zero would be contained in the credible interval of some coefficients. It is more informative to look at whether zero is contained within the credible interval for the entire smooth and therefore the spline can be said to be identically zero.

Plots of the posteriors of the marginal effects are a very informative way of both visualising splines and checking that the output of the model is consistent with the physical system that the statistical model is attempting to represent.

For plotting, we calculate a new basis with evenly spaced covariate values between the extreme values of each covariate in the data. This gives a basis with the same knots as the basis used for fitting the model but with a set of covariates that will give a visually more pleasing plot. The mean and 95\% credible regions for these posteriors are calculated by using the output from the Gibbs sampler. Bivariate splines will be plotted as surface and/or contour plots.

For a Generalised Linear Mixed Model, which encompasses a GAM with splines for random effects \citep{inlaglmm07},
\begin{equation*}
 \vec{y} = g^{-1} \left( \widetilde{\mat{X}}\widetilde{\vec{\beta}} + \widetilde{\mat{Z}}\widetilde{\vec{\gamma}} \right) + \vec{\varepsilon},
\end{equation*}
the fitted values are given by
\begin{align*}
\widehat{\vec{y}} =& \widetilde{\mat{X}}\widehat{\vec{\beta}} + \widetilde{\mat{Z}}\widehat{\vec{\gamma}} \\
                  =& {\mat{X}} \left( {\mat{X}}^T {\mat{X}} + \mat{\Lambda} \right)^{-1} {\mat{X}}^T \vec{y}
\end{align*}
where ${\mat{X}} = \left[ \widetilde{\mat{X}} \pipe \widetilde{\mat{Z}} \right] \in \mathbb{R}^{n \times (l + k)}$ (for $l$ parametric terms and $k$ non-parametric basis functions) and
\begin{equation}
 \mat{\Lambda} = \begin{bmatrix} \mat{0}_{l \times l} & \mat{0}_{l \times k} \\ \mat{0}_{k \times l} & \sigma^2 \left( \textnormal{cov} (\vec{\gamma}) \right)^{-1} \end{bmatrix}
\end{equation}
is the posterior precision of $\vec{\beta}$ from \ref{eq:beta} \citep{ruppertwandcarroll}.

The total effective number of degrees of freedom of the model is given by the trace of the hat matrix
\begin{equation}
\mat{H} = \left({\mat{X}}^T {\mat{X}} + \mat{\Lambda}\right)^{-1}{\mat{X}}^T {\mat{X}} \label{eq:df}
\end{equation}
and for each non-parametric term, the trace of the block matrix of the hat matrix corresponding to the appropriate coefficients gives the effective number of degrees of freedom for that non-parametric term \citep{inlaglmm07}. 


Rather than estimate the mean effective degrees of freedom \textit{post hoc} with the posterior means of the parameters we can use the samples from the MCMC chains as they are drawn to construct $\mat{H}$. In this way, we obtain the distribution of the number of parameters for each term by examining the density estimate of the trace of the blocks of $\mat{H}$.

The Deviance Information Criterion (DIC) is a Bayesian analogue of the Akaike Information Criterion and  penalises the deviance, a measure of the goodness of fit of a model, by the effective number of parameters in the model \citep{dic02}. The effective number of parameters can be compared to the effective degrees of freedom (described above). The effective number of parameters is based on the deviance and so the DIC does not require the integrating out of random effects parameters like Schwarz's Bayesian Criterion \citep{sbc78}.

While calculation of the credible interval for $\vec{\phi}$ will provide information about the degree of autocorrelation of the residuals, the autocorrelation function of both $\vec{\varepsilon}$ and $\vec{u}$ can be used to check how much autocorrelation remains. The posterior covariance matrix from the chains of $\vec{\varepsilon}$ and $\vec{u}$ can also be used to characterise the remaining autocorrelation of the residuals.

The probability integral transform (PIT) will be used to assess the quality of the model fit. The PIT is a measure of the cumulative density of the forecast values from their predictive distribution \citep*{dawid84, diebold98}. This predictive CDF, denoted $F_i$, is to be compared to an unknown ``true'' CDF $G_i$ through the observed values that the true physical process generates, $y_i$, and the forecast values, $\widehat{y}_i$. The ideal forecasting model is achieved when $F_i = G_i$. A necessary condition for choosing a forecasting model is that the distribution of $F_i (\widehat{y}_i)$ is uniform, corresponding to ideal forecasts.

The PIT for each forecast value is calculated during the forecasting step as
\[
F_i (\widehat{y}_i) = \mathrm{ecdf} \left( \widehat{y}_i; \, \widehat{\mu}_i + \widehat{\varepsilon}_i \right) 
\]
where $\mathrm{ecdf}$ is the empirical cumulative density function of the posterior samples of $\widehat{\mu}_i + \widehat{\varepsilon}_i$ from the forecasting routine. The PIT is visualised by plotting the histogram and autocorrelation of $F_i (\widehat{y}_i)$. If the PIT is uniformly distributed, then the values of the probability density function corresponding to the PIT's cumulative density function should be a normal distribution with mean 0 and standard deviation 1. The autocorrelation of the PIT should not be significant beyond the number of steps used in forecasting.

Many competing models may have a uniform PIT so the uniformity is not a sufficient condition for choosing one model over another \citep{hamill01}. To remedy this, \citet{gneiting07jrssb} recommend maximising the sharpness of the predictive distribution, i.e. choosing forecasts which are highly concentrated about the observed values. The variance of the forecast values will be estimated by characterising their uncertainty with the width of the 95\% credible interval of the simulated forecasts (which are distributed normally). The 95\% credible interval corresponds to 1.96 standard deviations of the estimate, so dividing the interval half width by 1.96 and taking the square will give an estimate of $\textnormal{Var}\left( \widehat{\vec{y}} \right)$.

\section{Case studies}\label{sec:case}
To demonstrate the performance of the methodology in terms of its ability to fit smooth covariate effects and autocorrelated residuals we provide two case studies. The first is a simulation study with a univariate cyclic smooth and a bivariate interaction term. The second case study involves modelling and forecasting ultrafine particle number concentration in Helsinki, Finland, a real world data set that exhibits a significant amount of autocorrelation.

\subsection{Simulation}\label{sec:sim}
As a first case study, to illustrate the method, we provide an example with simulated data using a Gaussian likelihood, as in \citet{Chib93}. Data is simulated as described in \ref{eq:sim}. Two covariates, $x$ and $y$, are each drawn from a uniform distribution and a non-linear interaction term constructed that requires that a 2D spline be fit. A sinusoidal function to simulate a temporal trend (at evenly spaced intervals) is added, as is some autocorrelated noise.

\begin{equation}
\begin{aligned}
\vec{x}, \vec{y} \sim & \, \mathcal{U}\left(0,1 \right) \\
\vec{t} = & \, 1, 2, \ldots, 24 \\
\mu_i = \, & \sin\left(\pi x_i \right) \left( 1-x_i y_i^2 \right) + \frac{\sin\left( \frac{2\pi t_i}{24} \right)}{2} \\
\left(1 + 0.4 L \right) \varepsilon_i \sim & \, \mathcal{N}\left( 0, 0.1 \right)
\end{aligned}\label{eq:sim}
\end{equation}

The model for this simulation thus contains: a univariate spline for $\vec{t}$ with a basis of six second order cyclic B-splines with a second order penalty prior; a bivariate spline for fitting $\vec{x}$ and $\vec{y}$ together, consisting of the tensor product of two second order non-cyclic B-splines bases with six basis splines each and a second order penalty prior for each direction; an intercept term with a weakly informative prior; and an AR(1) model for the residuals.

\subsection{PNC in Helsinki}\label{sec:csfinland}
Ultrafine particles are of interest in a range of areas, including physics \citep{hinds1999}, urban planning \citep{healthybuildingsbook98} and epidemiology \citep{dehartog03,heireport,hei2010}. In the urban environment it has been established that a significant portion of the ultrafine particles come from vehicles \citep{ambient08,virtanen06,pohjola07,hussein07} and that the particle number concentration (PNC) varies non-linearly and irregularly over the day, week and with meteorology \citep{jaime5year, weeklytrends2002, hussein06, ketzel2004, perez2010, wu2008, wehnerWiedensohler2003}. 

Continuously measured time series of ultrafine PNC exhibit temporal trends, dependence on meteorology and autocorrelation \citep{wehnerWiedensohler2003, perez2010, wu2008, hussein04, hussein06, jarvi09}. As such, the desire for flexible models which take these features into account without specifying the functional form of the relationship has motivated the use of the Generalised Additive Model with splines as basis functions and Generalised Linear Models with Fourier series basis functions.

In this case study we use hourly averaged size fractionated PNC, recorded at the SMEAR-III station at the Kumpula campus of the University of Helsinki using a twin DMPS system \citep{jarvi09}. Meteorological data was recorded at the university campus on the rooftop of the Physicum building. For further information on the data collection, see sections 2.2-2.4 of \citet{molgaard2011}.

Four different specifications of the model are fitted and model choice is made with the DIC. The reduction in autocorrelation is analysed and estimates of the fitted splines are provided.

\begin{figure}[ht]
\centering
\includegraphics[width=\linewidth]{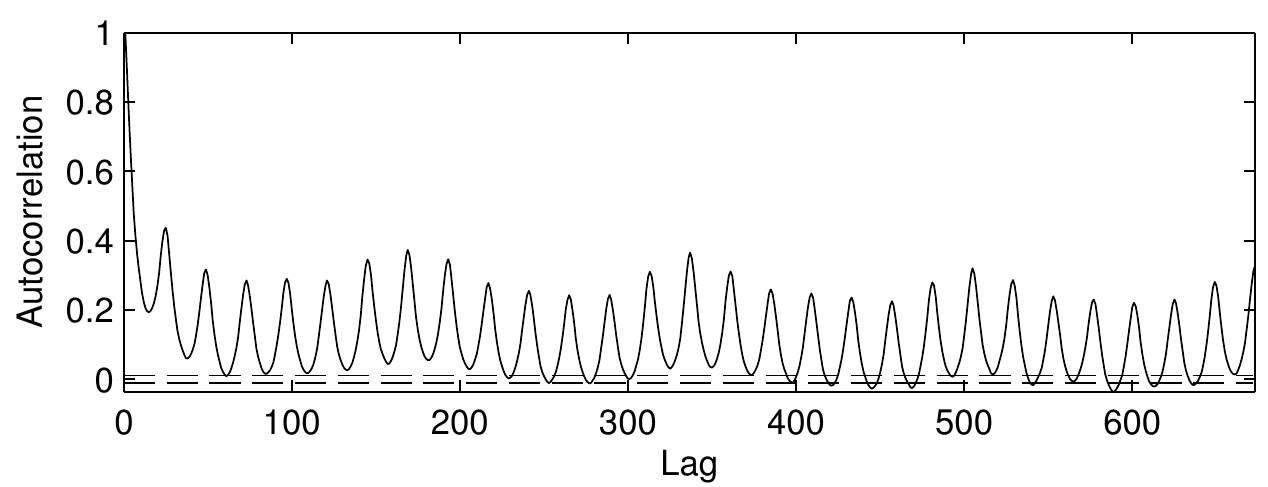}
\caption{Autocorrelation function of ultrafine PNC in Helsinki for the data described in Section \ref{sec:csfinland}. Local maxima occurring at lags which are multiples of 24 correspond to a daily trend relationship, multiples of 168 to a weekly relationship. The maximum lag here corresponds to four weeks. The dashed lines represent the 5\% significance level.}
\label{fig:acf}
\end{figure}

The autocorrelation function of the ultrafine PNC in Helsinki is shown in Figure \ref{fig:acf}. This autocorrelation motivates the use of splines to estimate temporal trends and a model for the residuals which can capture any remaining variation.

\begin{table*}[ht]
\centering
\begin{tabularx}{\linewidth}{Xlccl}
\toprule
Parameter(s) & Basis type & Basis size(s) & Periodic & Covariance structure \\
\midrule
Hour of the day & B-spline & 6 & Y & 2nd order penalty \\
Day of the week & Factor & 7 & N & iid \\
Day of the year & B-spline & 6 & Y & 2nd order penalty \\
Wind direction & B-spline & 8 & Y & 2nd order penalty \\
Wind speed & Thin plate & 6 & N & iid \\
Temperature & B-spline & 8 & N & 2nd order penalty \\
Traffic & B-spline & 8 & N & 2nd order penalty \\
Relative humidity & B-spline & 8 & N & 2nd order penalty \\
Solar radiation & B-spline & 8 & N & 2nd order penalty \\
\bottomrule
\end{tabularx}
\caption{List of covariates and their univariate bases used in fitting a semi-parametric regression and forecasting model to the Helsinki data.}
\label{tab:hel}
\end{table*}

Table \ref{tab:hel} describes the basis functions used in fitting the regression model. The thin plate spline used for wind speed includes a linear fixed effect and the random effects are five first order polynomials, $|x_i - \kappa_k|$. A tensor product of the wind speed and wind directions is used, with a basis size of 48. An attempt was made to use P-splines for the wind speed term but the presence of splines with no support at the extremes of the covariate space is undesirable.

Four models are fit and the ``best'' model is chosen with the DIC, calculated at the final stage of the model fitting. The first model is the one described by Table \ref{tab:hel} and the joint wind speed and wind direction term described above. The second model replaces the univariate splines for temperature, traffic and relative humidity with a tensor product of each of those covariates with wind direction. The third model retains the univariate splines as well as the tensor products. The fourth model is the second model with the daily trend replaced with a tensor product of the daily trend and annual trend in order to recognise that the daily trend may change over the year.

The annual trend was excluded from the first three models as solar radiation and temperature exhibit very strong annual trends \citep{fmi1981-2010}. Its inclusion in the fourth model is for the purposes of the tensor product with the daily trend. We will provide a plot of the fitted tensor as well as the marginal annual trend and marginal daily trend. These marginal trends are obtained by averaging over a prediction basis defined on a mesh consisting of one copy of each unique combination of time of day and day of the year.

For the autocorrelated residuals, lags at values of 1 hour, 24 hours and 168 hours are used. These lags represent, respectively, an attempt to capture the leftover hour to hour variation from fitting the daily trend, the leftover daily variation from fitting the daily trend and the leftover weekly variation after fitting the day of the week factor term.

The models will be fit to the first three years of the four years of data and predictions, including the autocorrelated residuals, will be made for the following year with the measured covariate values.

\section{Results and Discussion}\label{sec:results}
\subsection{Simulation}\label{sec:simfit}
MCMC estimation was conducted by drawing 5000 MCMC samples from the posterior; 500 initial samples are discarded as burn in.

\begin{figure*}[htbp]
\centering
\includegraphics[width=0.8\textwidth]{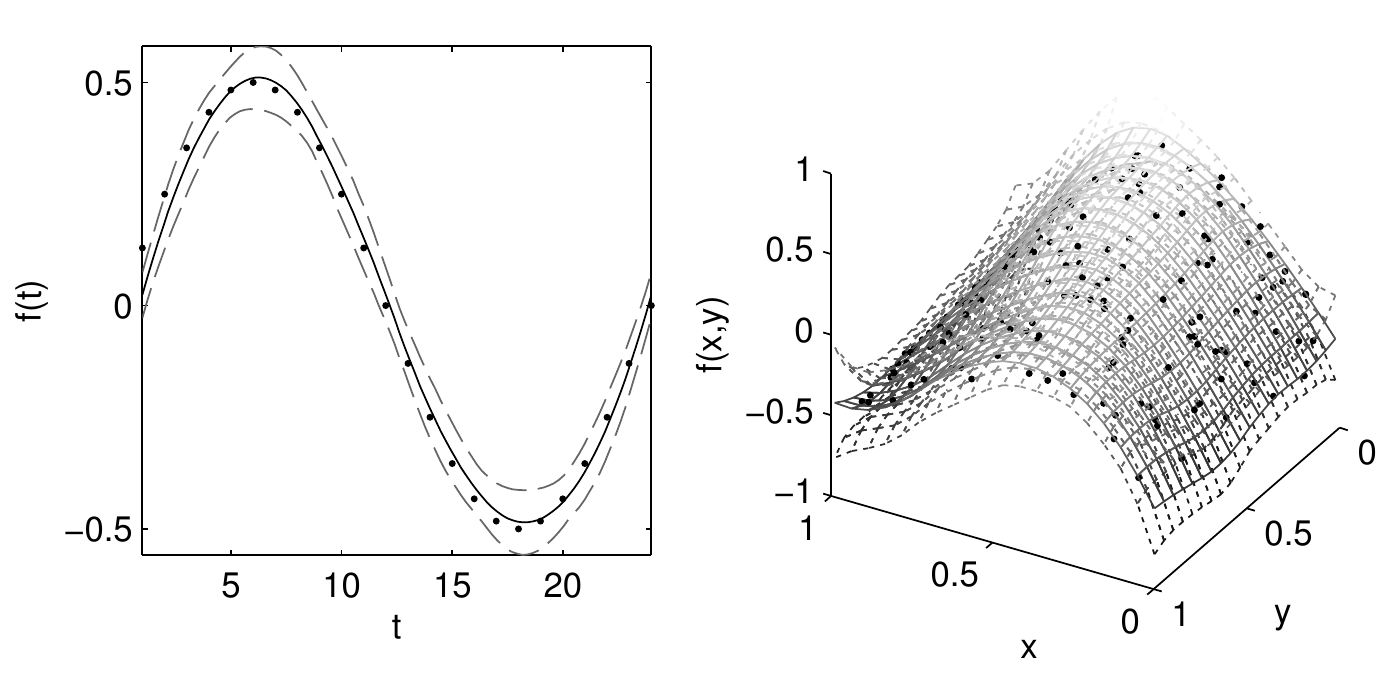}
\caption{Marginal effects of splines (solid), 95\% credible interval (dashed) and the true values from the simulated data (dots).}
\label{fig:splines}
\end{figure*}

We see (Figure \ref{fig:splines}) that the periodic term is estimated accurately by the model, as is the non-linear interaction. We note that the credible intervals for the periodic term are quite similar in width across the entire covariate domain. This is due to the periodicity of the basis, uniform spacing of the covariate values and that each unique covariate value occurs the same number of times.

The fitted two dimensional spline captures the curvature and asymmetry of the simulation function (Figure \ref{fig:splines}).

\begin{sidewaysfigure*}[htpb]
\centering
\includegraphics[width=\linewidth]{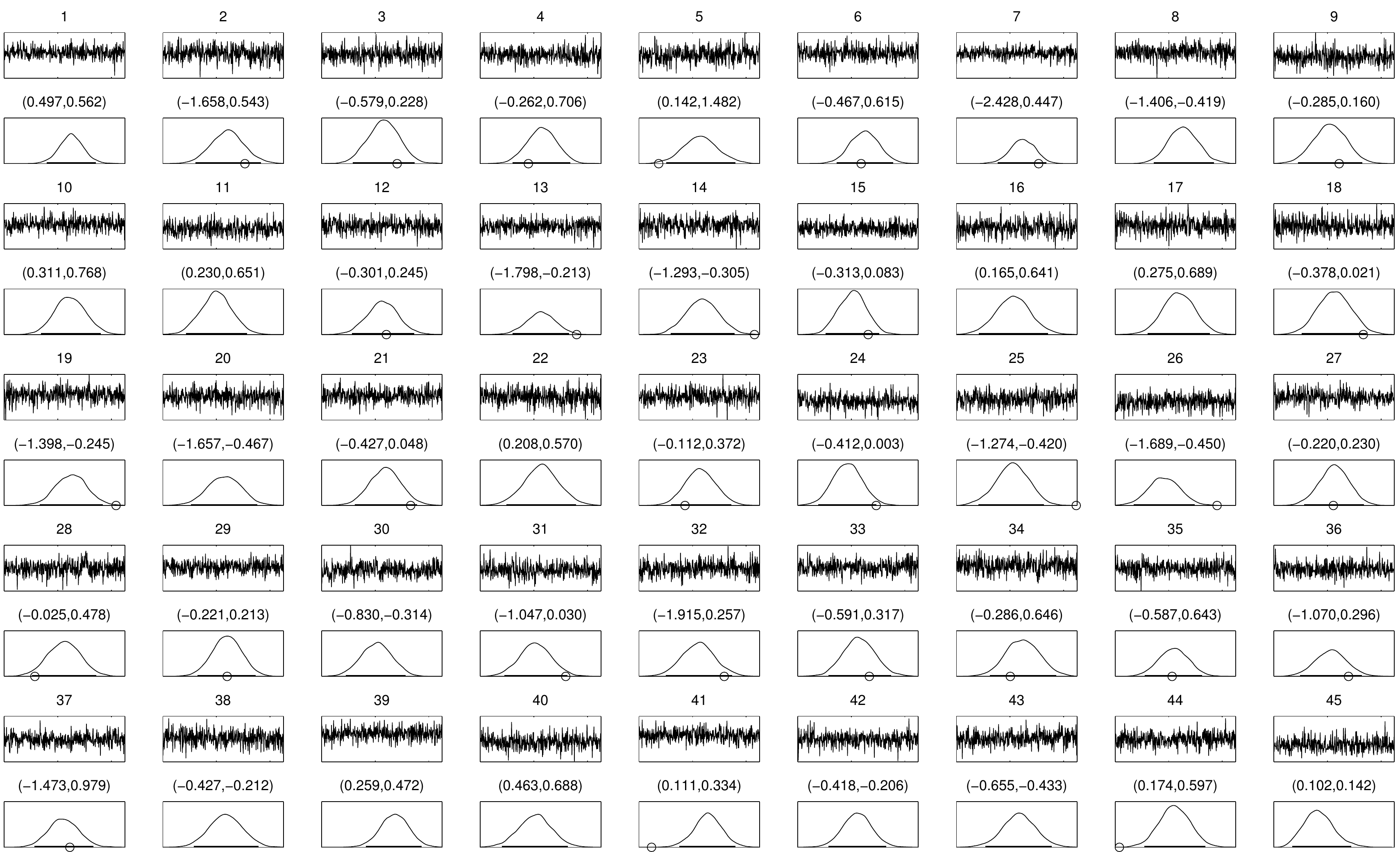}
\caption{MCMC posterior summaries. The 95\% credible interval is represented by a horizontal black bar in the density plot and the quantiles are given to three decimal places. Where zero is within the range of the posterior samples, it is plotted on the density plot as a circle.}
\label{fig:traces}
\end{sidewaysfigure*}

Traces and densities of the posterior samples drawn from the posteriors of all parameters have converged and are unimodal and normally distributed except for the standard deviation parameter whose square is inverse Gamma (Figure \ref{fig:traces}).

\begin{figure}[htbp]
\centering
\includegraphics[width=\linewidth]{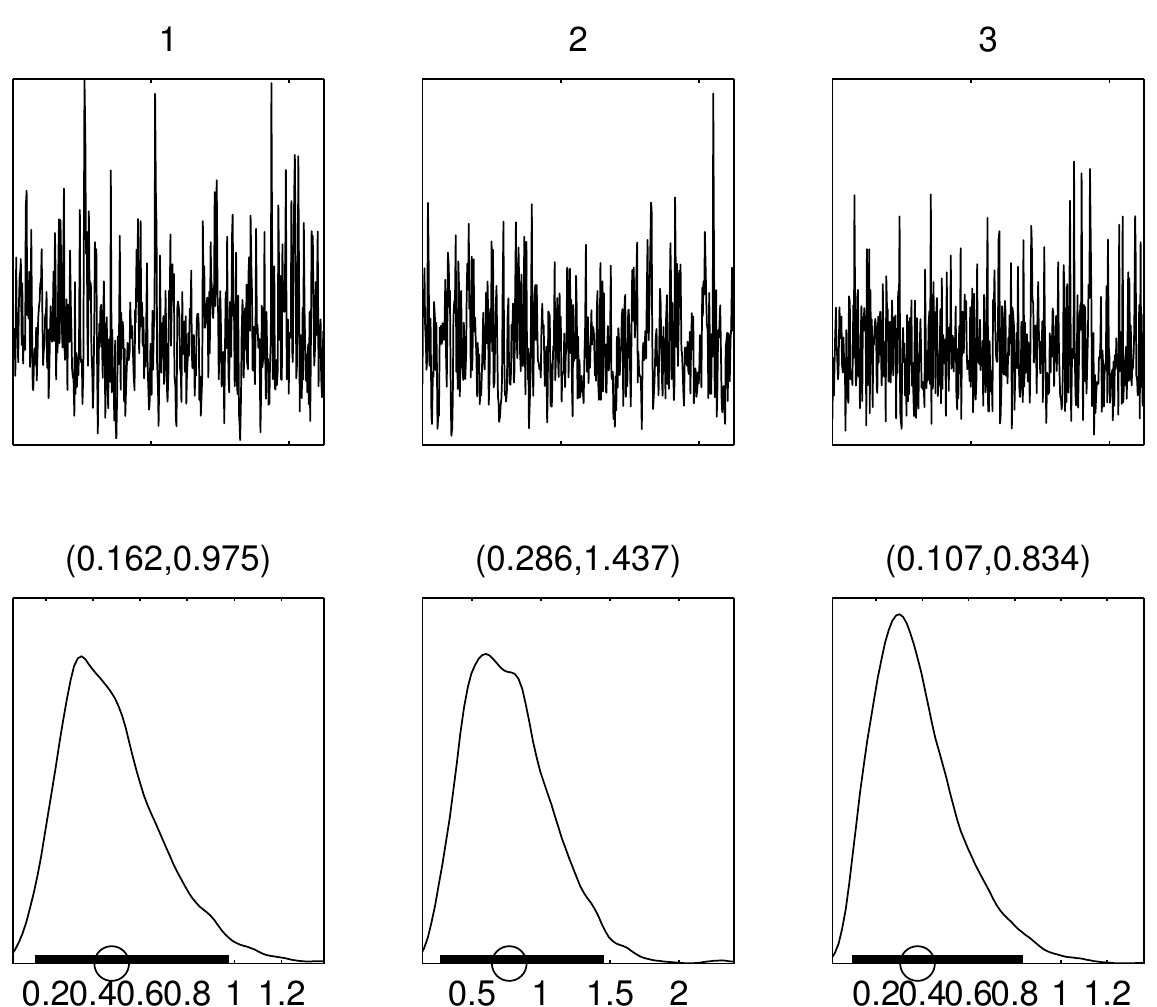}
\caption{Trace and density plots for MCMC samples of smoothing parameters for simulated data. From left, $\lambda_x$, $\lambda_y$ and $\lambda_t$.}
\label{fig:lambda}
\end{figure}

Figure \ref{fig:lambda} gives trace and density plots for the smoothing parameters, which are Gamma distributed. We see that $\lambda_x$ and $\lambda_y$ have different, though not overly so, credible intervals. Had the 2D covariate been more oscillatory it is likely that we would end up with markedly distinct smoothing parameters \citep[see][sec. 6]{eilersmarx2003}. The priors for these smoothing parameters have a maximum at zero. The maximum in each of these posteriors now occurs at a non-zero value. While the parameters are still (approximately) Gamma distributed, they are no longer $\Gamma(1,\cdot)$.

The first parameter is the intercept term, $\beta_0$, parameters 2 to the 37 are the coefficients of the two dimensional spline, parameters 38 to 43 are the coefficients of the periodic spline. Parameter 44 is $\phi$, the autoregressive parameter from the AR(1) model for the residuals and the final parameter is the standard deviation of the independent identically distributed errors, whose square is inverse gamma. All other parameters are marginally normally distributed.

\begin{figure}[htbp]
	\centering
\subfloat[True 2D covariate]{\includegraphics[width=0.8\linewidth]{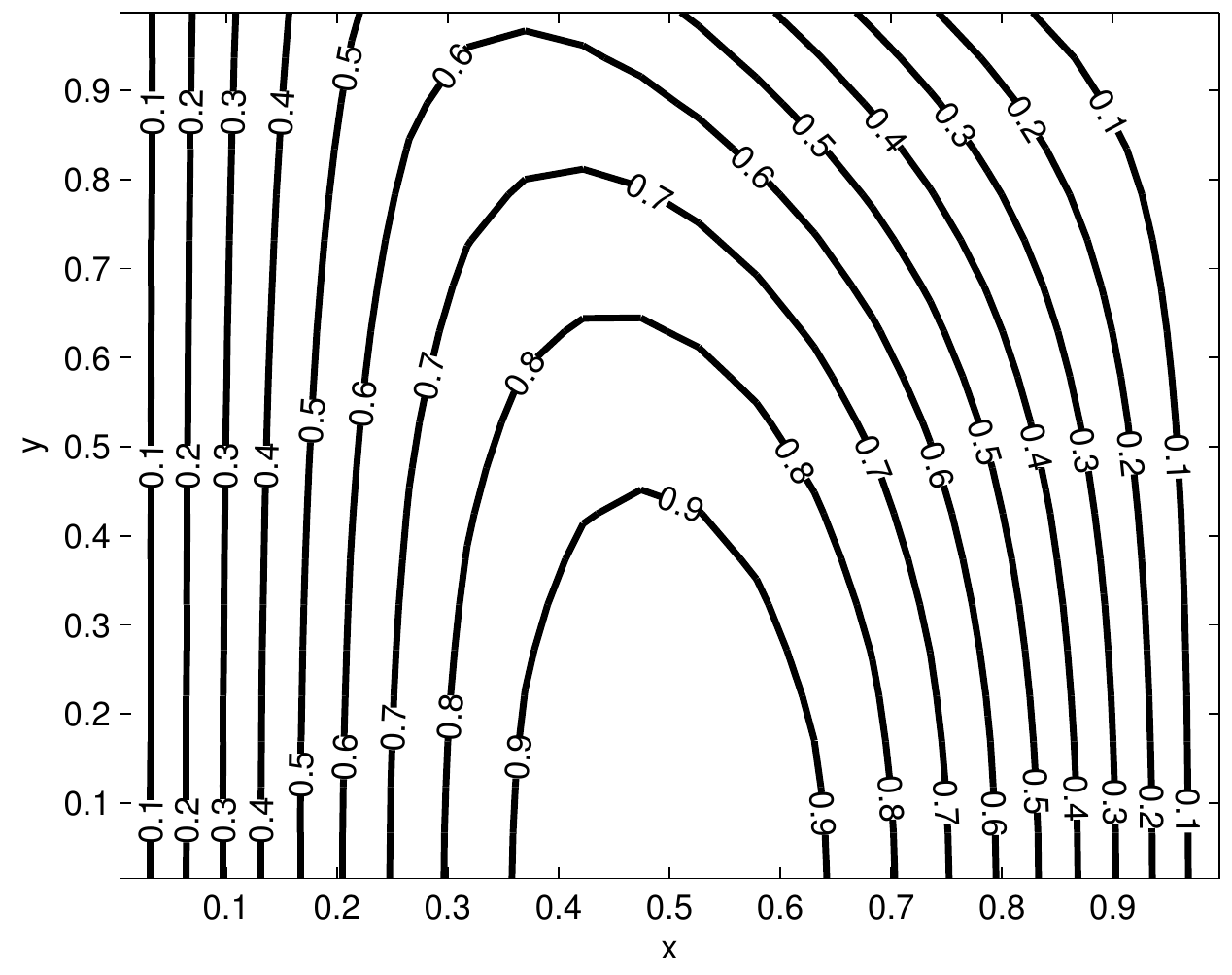}} \\
\subfloat[Combined partial effects of 2D spline and constant]{\includegraphics[width=0.8\linewidth]{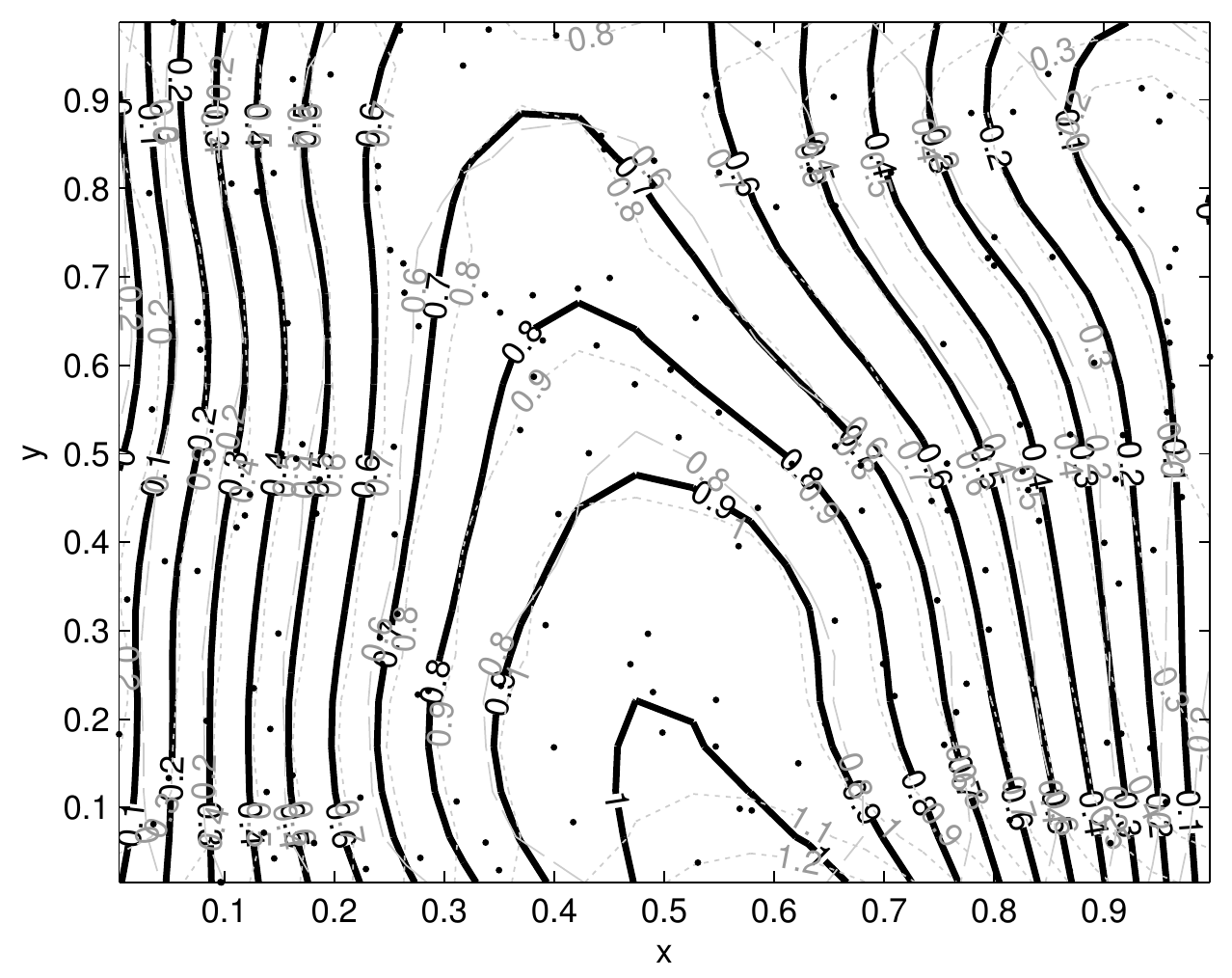}}
	\caption{Comparison of contours of the simulated 2D covariate from \ref{eq:sim} and the fitted 2D spline. Observed locations of $x$ and $y$ are plotted as points. The 2.5\% quantile is represented with long grey dashes, the 97.5\% quantile is represented with short grey dashes.}
	\label{fig:spatz}
\end{figure}

The contours of the fitted 2D spline show that the simulated 2D covariate effect has been accurately reconstructed (Figure \ref{fig:spatz}). The estimate for $\beta_0$ has been added to the 2D spline to allow for a more direct comparison of the values of the contours, as the fitted spline is centred about zero but the 2D covariate effect is not. Note that both the simulated temporal trend and its corresponding univariate spline are centred about zero and the AR errors also have a zero mean. The credible intervals for the two dimensional spline are quite wide in regions where there are not many observations and at the edges of the covariate space. The mean surface of the 2D smooth has no local maxima or minima, the presence of which would indicate excessive wiggliness.

\begin{figure}[htbp]
\centering
\subfloat[Covariance of posterior samples of autocorrelated errors, $\varepsilon_i$, of each observation]{\includegraphics[height=0.4\linewidth]{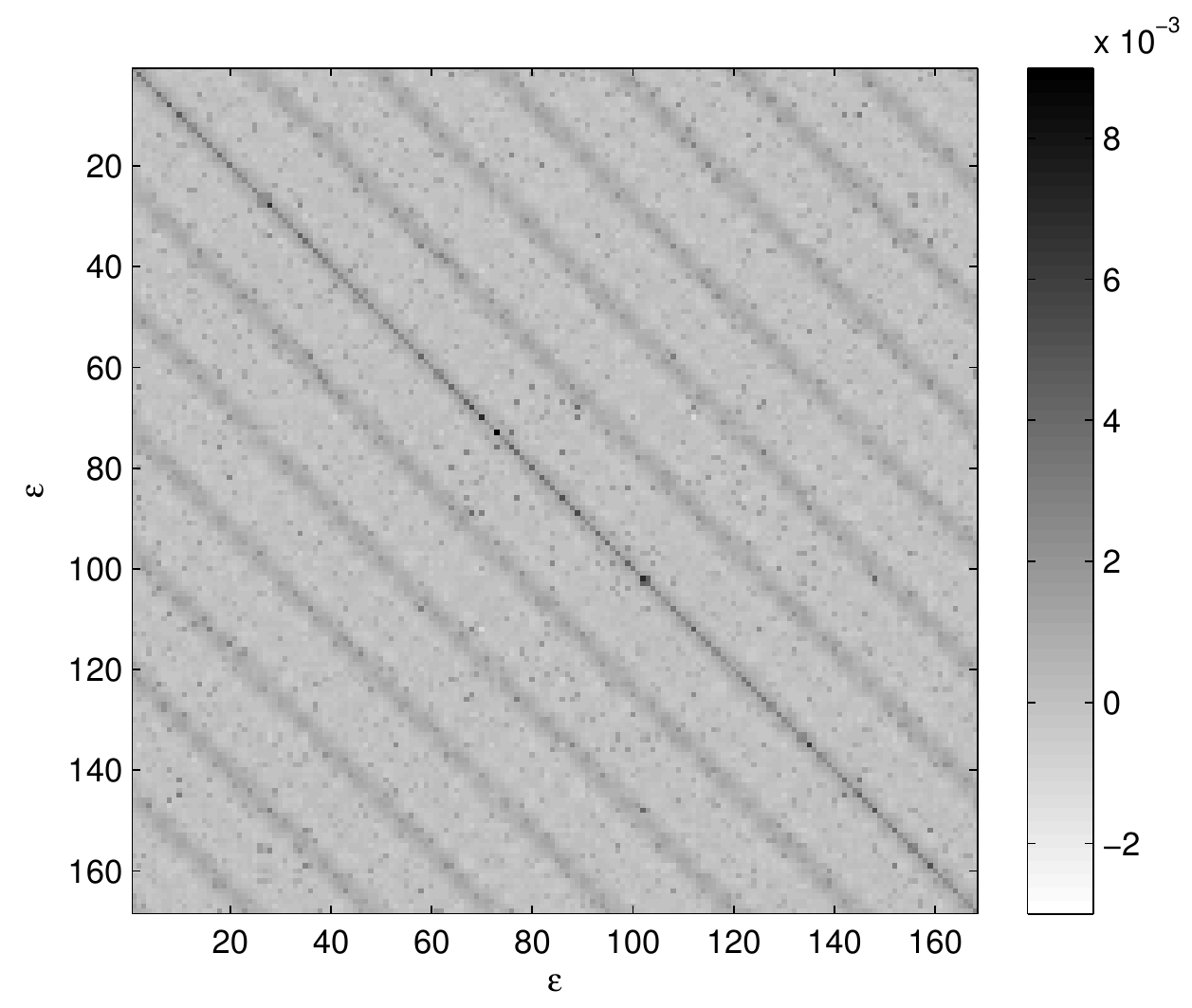}}
\quad
\subfloat[Covariance of posterior samples of iid errors, $u_i$, of each observation]{\includegraphics[height=0.4\linewidth]{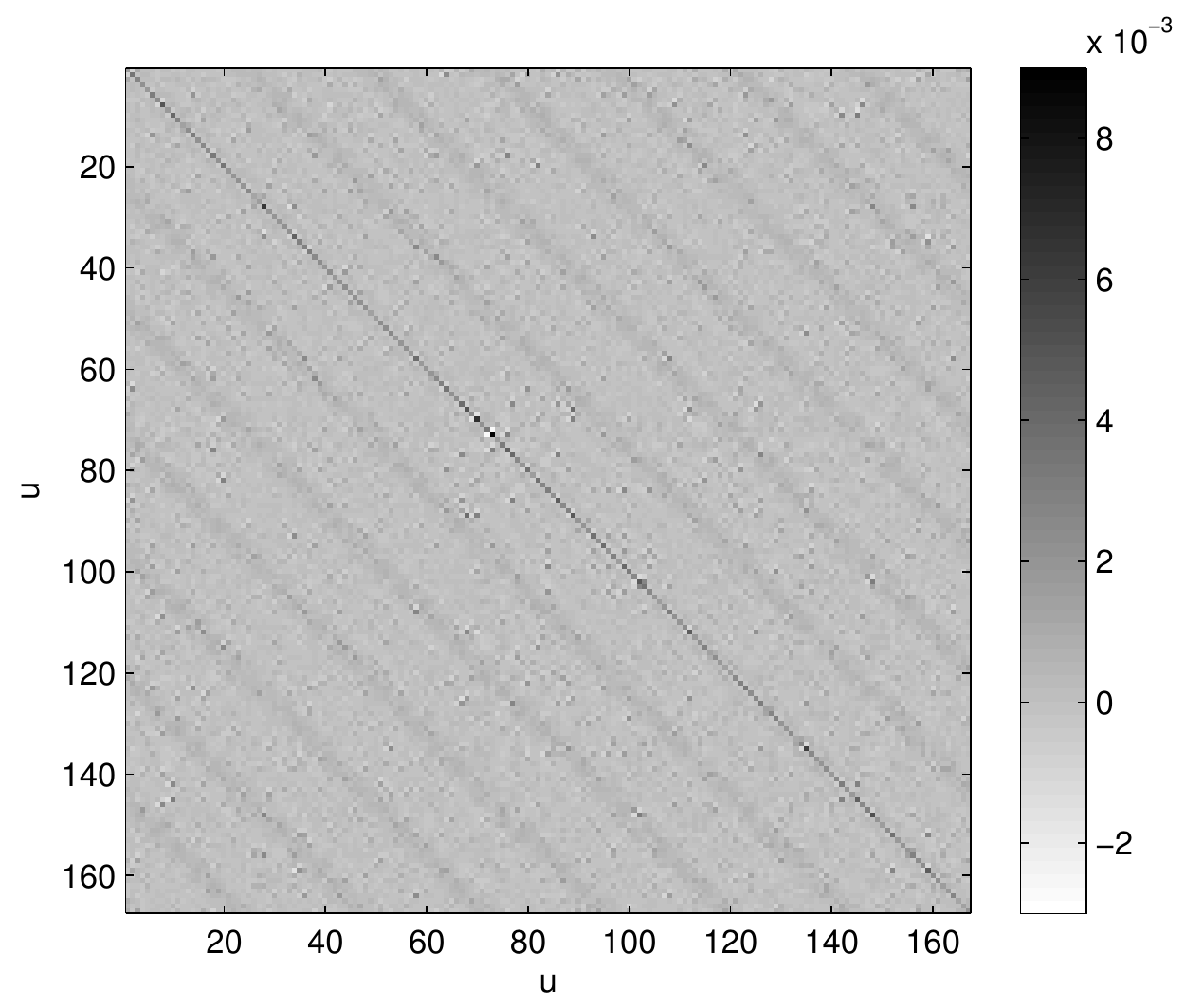}}
\caption{Covariance matrices of posterior samples of autocorrelated errors, $\vec{\varepsilon}$ and independent identically distributed residuals $\vec{u}$, from simulated model.}
\label{fig:simerrors}
\end{figure}

Figure \ref{fig:simerrors} shows that the residuals in $\vec{u}$ have smaller autocovariance than the sampled values of $\vec{\varepsilon}$. The diagonal bands correspond to 24 lags, indicating that the variation 24 observations apart may be modelled by expanding the model for the AR residuals to contain lags 1 and 24. Even so, the lag 24 bands and the background are lighter for $\vec{u}$ than for $\vec{\varepsilon}$ suggesting that modelling the autoregressive nature of the residuals has reduced the correlation of the posterior samples of the residuals.

The effective degrees of freedom of the intercept are concentrated around 1 with very little variance; it would be troubling were this not the case (Table \ref{tab:simedf}). The 2D spline has between 21 and 27 effective degrees of freedom; the basis for this term has 36 elements, so each basis element requires, on average, fewer degrees freedom than would be required by the corresponding polynomial basis. Similarly, the size of the basis for the temporal trend is six basis elements (and would have been eight had we not insisted on periodicity of the basis) and the effective degrees of freedom of this term is slightly less than five.

\begin{table}[ht]
\centering
\begin{tabularx}{\linewidth}{X*{5}{r}}
\toprule
Term & 2.5\%  & 50\%   & 97.5\% & Mean   & SD \\
\midrule
Intercept & 1.00  &  1.00   & 1.00  &  1.00 &   0.00 \\
2D &  23.34 &  25.61 &  27.45 &  25.57  &  1.03 \\
Temporal &  4.95  &  4.98  &  4.99  &  4.98  &  0.01 \\
\midrule
Total &  29.32 &  31.59  & 33.43  & 31.55  &  1.03 \\
\bottomrule
\end{tabularx}
\caption{Effective degrees of freedom summary statistics for model fit to simulated data.}
\label{tab:simedf}
\end{table}

\subsection{PNC in Helsinki}\label{sec:finfit}
\newcommand{\samref}[1]{\subref{#1}}
All but the first year is forecast both one day and two days in advance. The forecast values are stored throughout the iterative forecasting such that all bar the first year of the data is forecast conditioned on the preceding data.

\begin{figure}[ht]
\centering
\subfloat[Probability integral transforms ]{\includegraphics[width=0.8\linewidth]{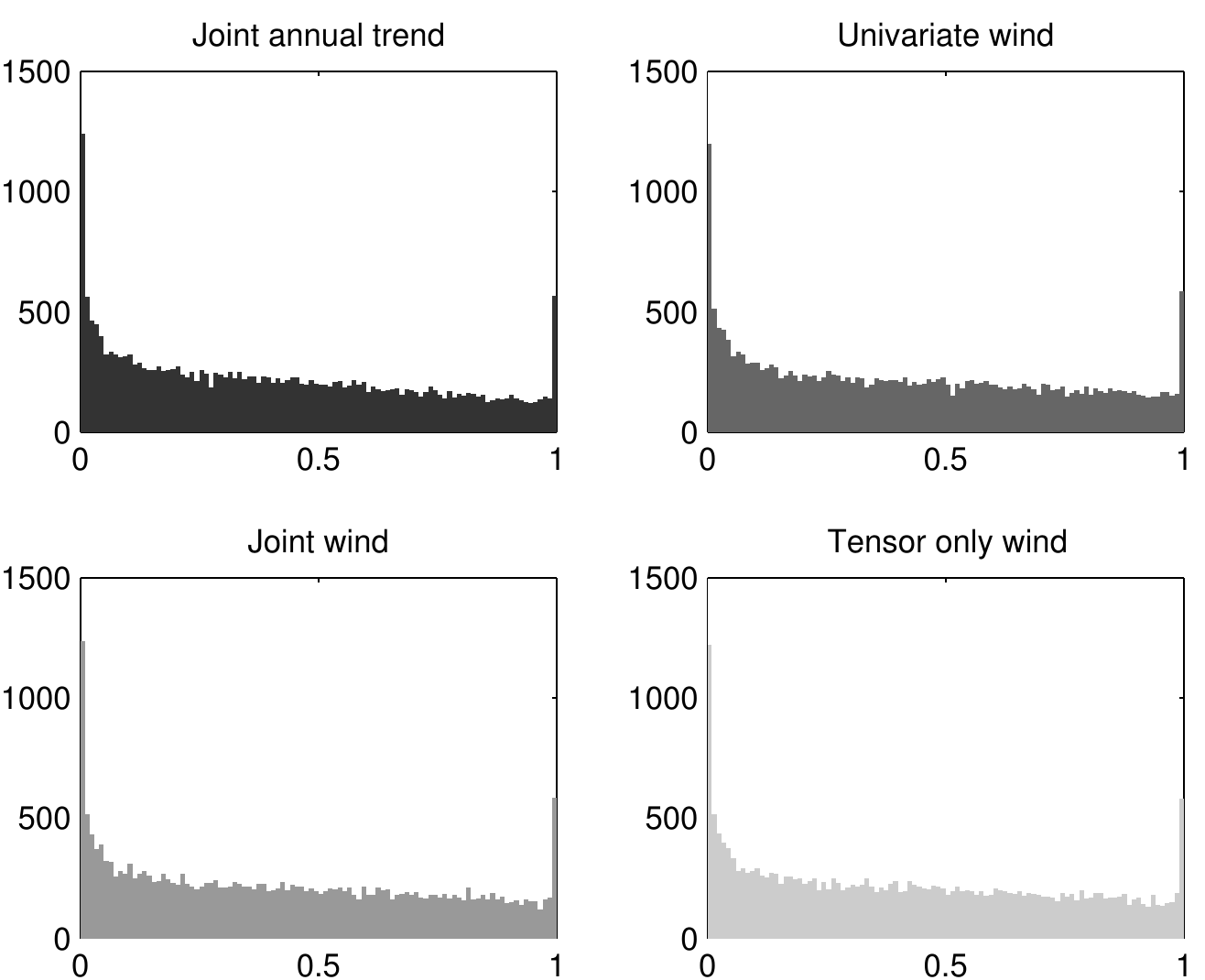}\label{fig:pithist}} \\
\subfloat[Density estimates of the variance of the forecast values.]{\includegraphics[width=0.8\linewidth]{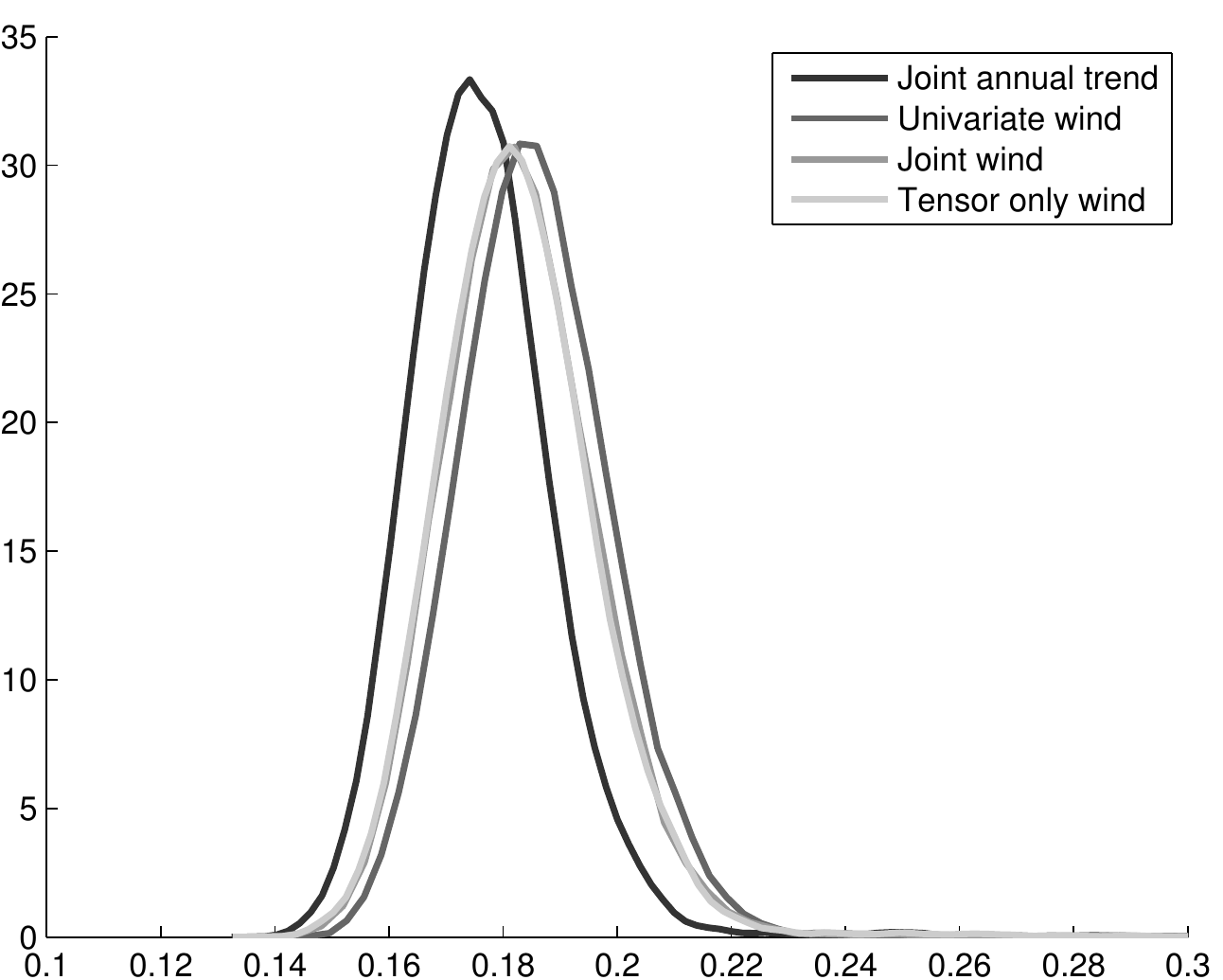}\label{fig:vars}}
\caption{PITs and density estimates of forecast values for each of the four models fitted.}
\label{fig:pit}
\end{figure}

Of the four models fitted, the model with the joint annual-daily trend had the lowest DIC (Table \ref{tab:finmodels}). The triangle shape of the PIT is very similar for each of the models and indicates biasedness in the samples towards slight underprediction (Figure \ref{fig:pithist}). The estimates of the variance of the modelled values (Figure \ref{fig:vars}) indicates that while the models provide estimates with similar variances, the joint annual-daily temporal trend with tensor product meteorology provides more concentrated forecasts. This model also has the lowest DIC out of the four models fitted. Therefore, this model represents the most efficient fit in terms of goodness of fit versus model complexity and has the most concentrated forecast values out of a group of models which perform similarly under the PIT. All subsequent analysis in this section is performed on this model.

\begin{table*}[ht]
\centering
\begin{tabularx}{\linewidth}{X*{4}{r}}
\toprule
Model & DIC & $p_D$ & $\overline{\textnormal{edf}}$ & $\mathrm{rank}(\mat{X})$ \\
\midrule
Univariate meteorology & 13023 & 96.809 & 86.7699 & 94  \\ 
Tensor meteorology & 12404 & 215.022 & 203.169 & 262  \\ 
Combination of univariate and tensor & 12406 & 218.360 &  205.405 & 286  \\ 
Tensor meteorology with annual-daily trend & 12055 & 244.166 & 230.318 & 292  \\
\bottomrule
\end{tabularx}
\caption{Summaries of fitting the models with differing treatments of meteorological covariates with respect to wind direction and temporal trends. DIC, effective number of parameters, mean effective degrees of freedom, and number of columns in the design matrix, $\mat{X}$.}
\label{tab:finmodels}
\end{table*}

\begin{figure*}[htbp]
\centering
\subfloat[Intercept]{\includegraphics[width=0.3\textwidth]{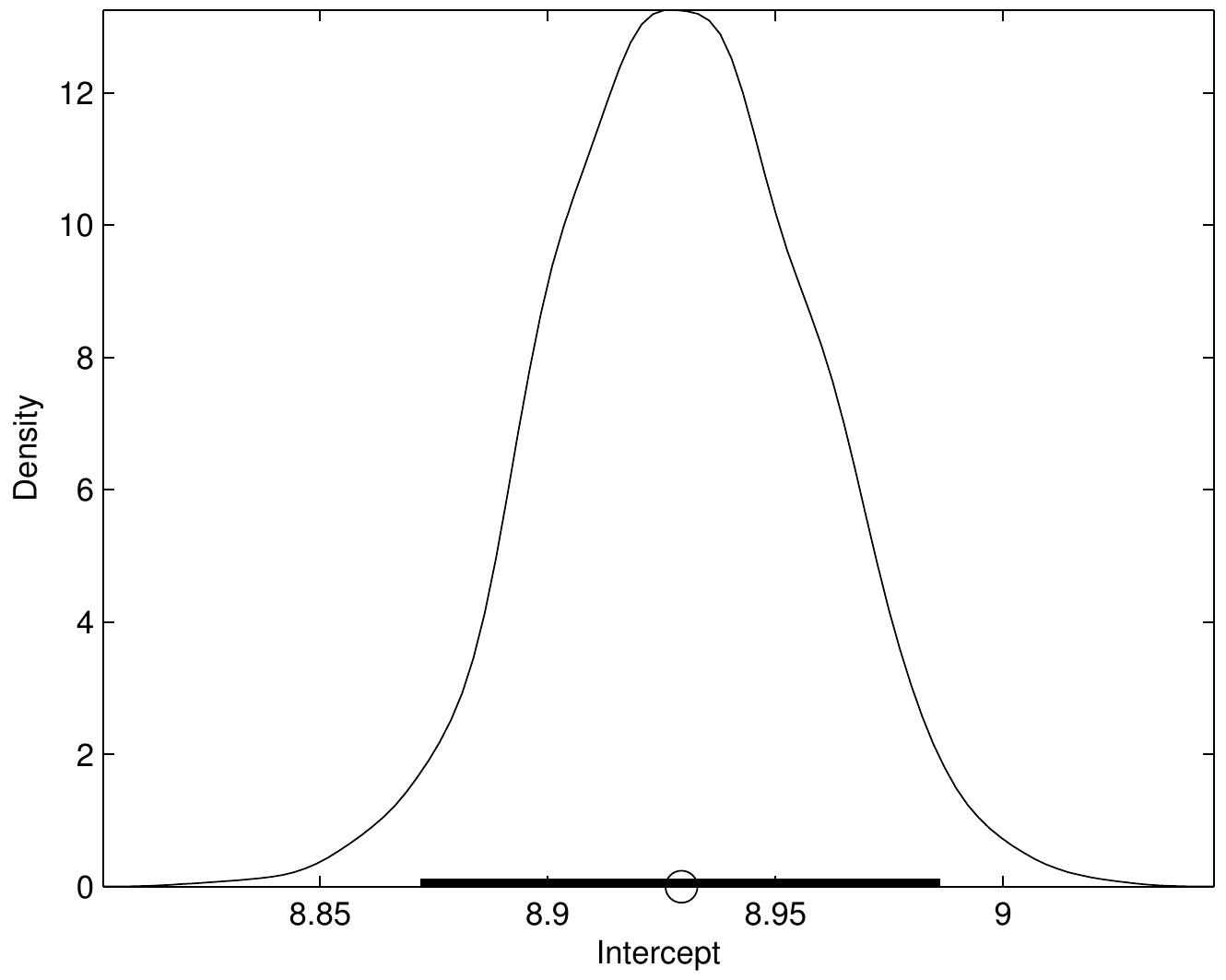}\label{fig:hbeta0}}
\subfloat[Standard deviation]{\includegraphics[width=0.3\textwidth]{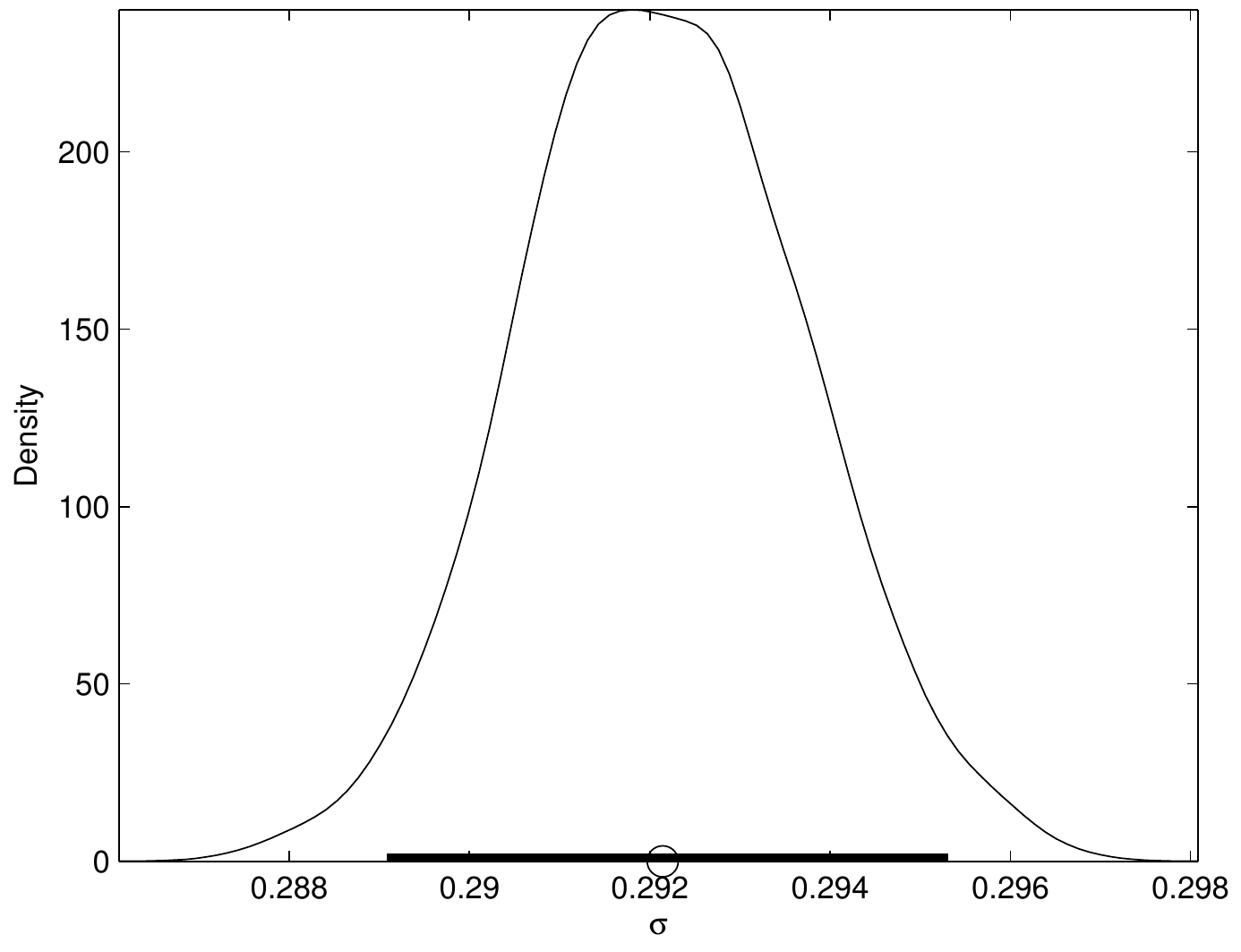}\label{fig:hsigma}} 
\subfloat[Joint daily and annual trend]{\includegraphics[width=0.3\textwidth]{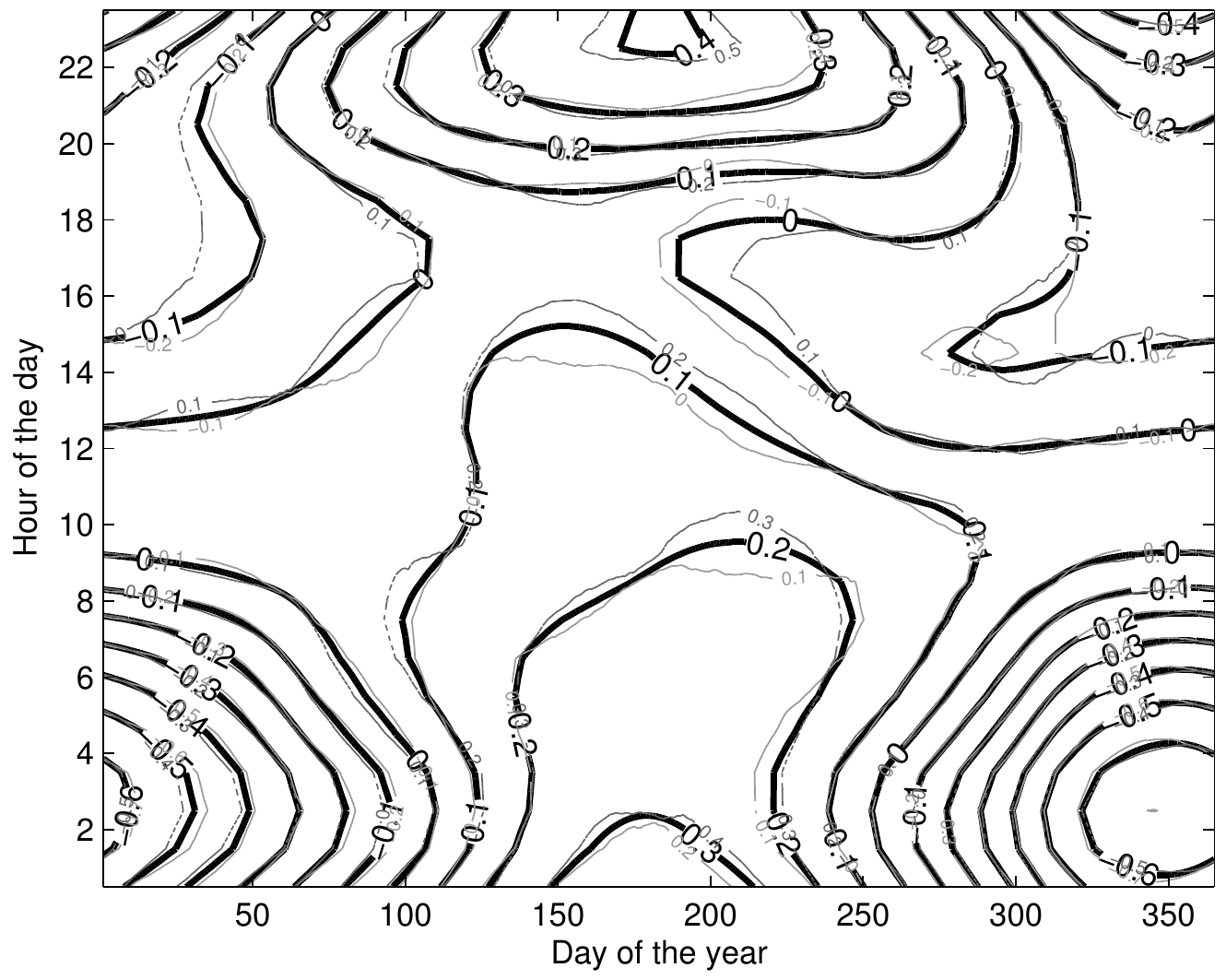}\label{fig:htime}} 
\\
\subfloat[Weekly trend]{\includegraphics[width=0.3\textwidth]{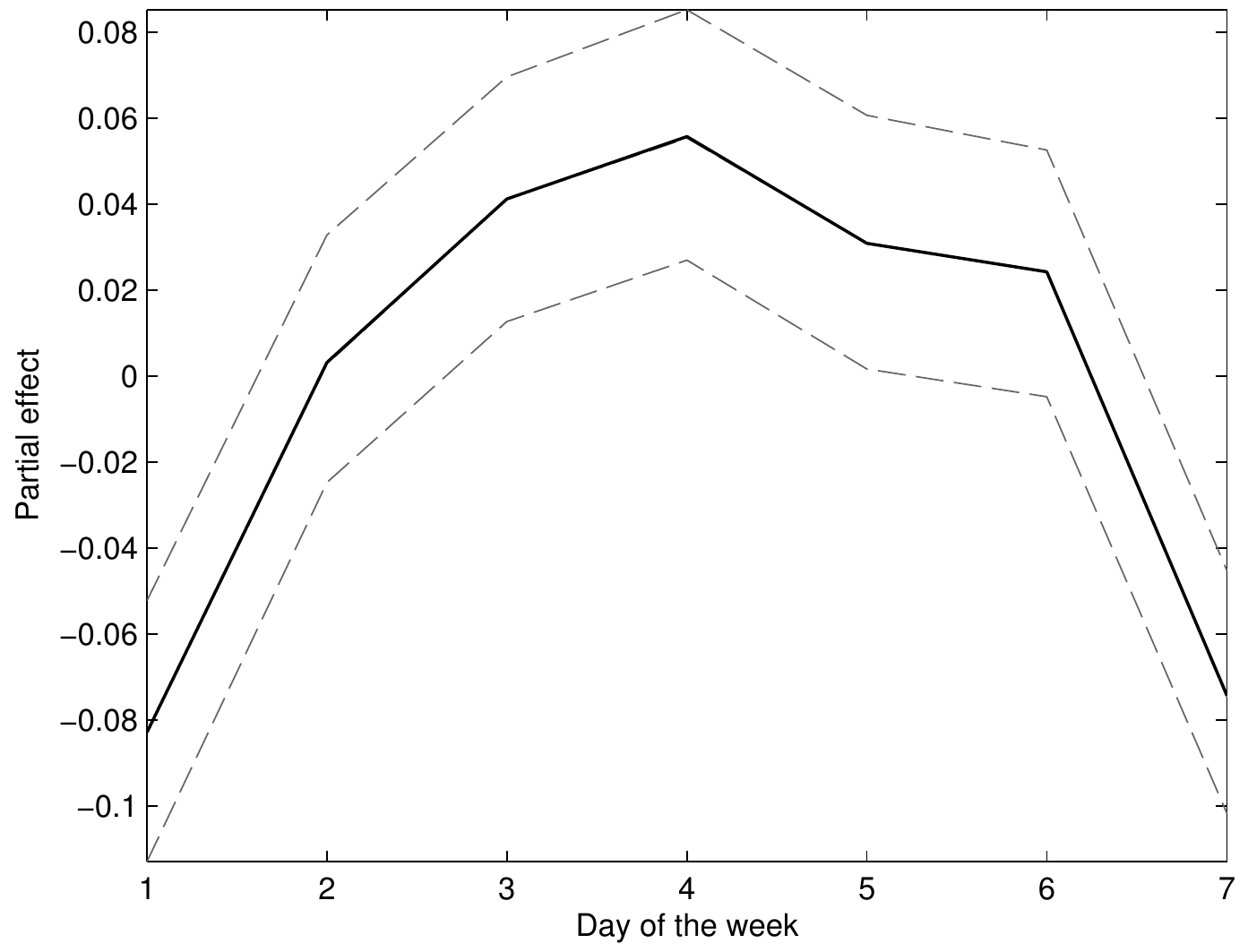}\label{fig:hdow}} 
\subfloat[Joint wind effect]{\includegraphics[width=0.3\textwidth]{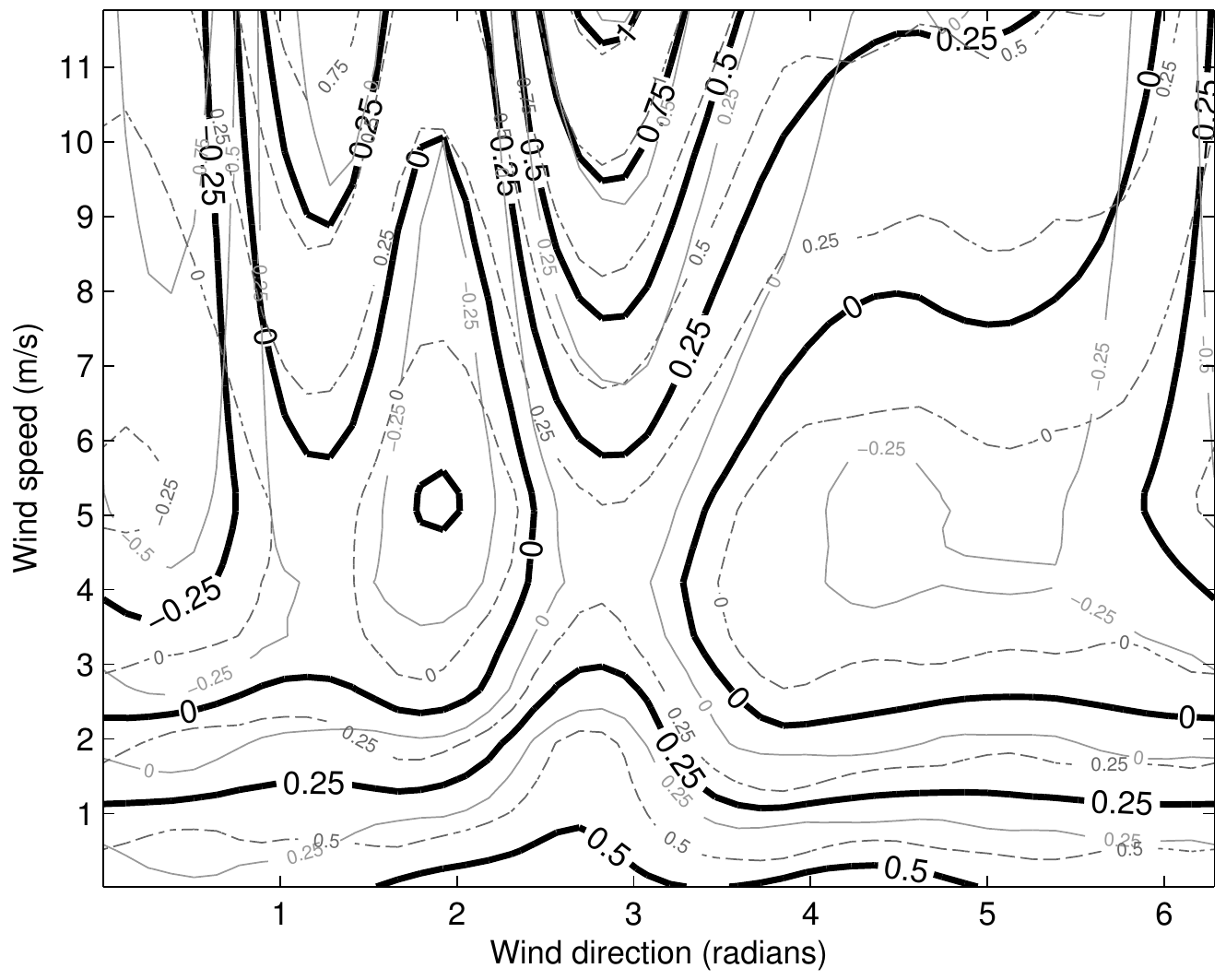}\label{fig:hwind}} 
\subfloat[Humidity]{\includegraphics[width=0.3\textwidth]{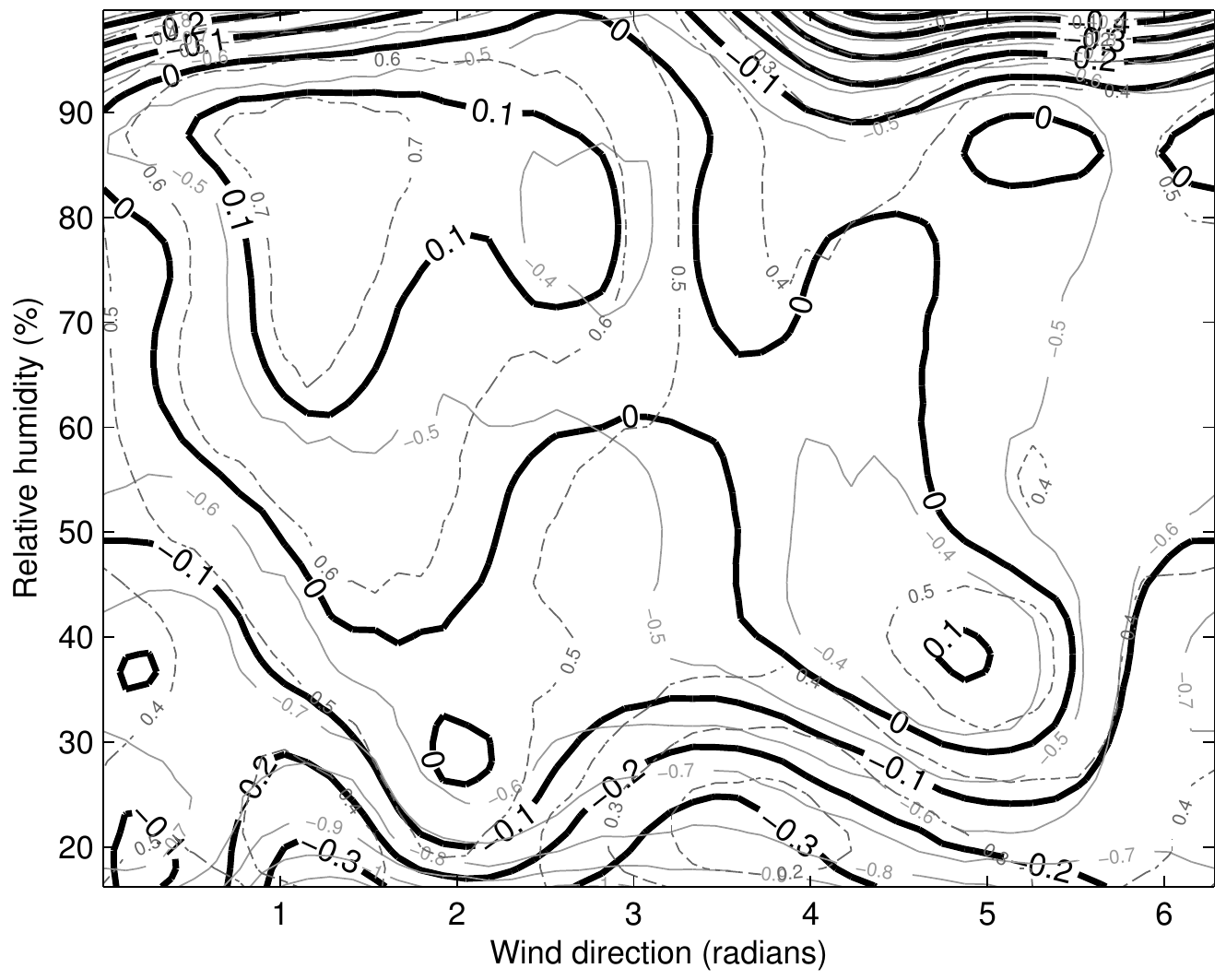}\label{fig:hrh}} 
\\
\subfloat[Temperature]{\includegraphics[width=0.3\textwidth]{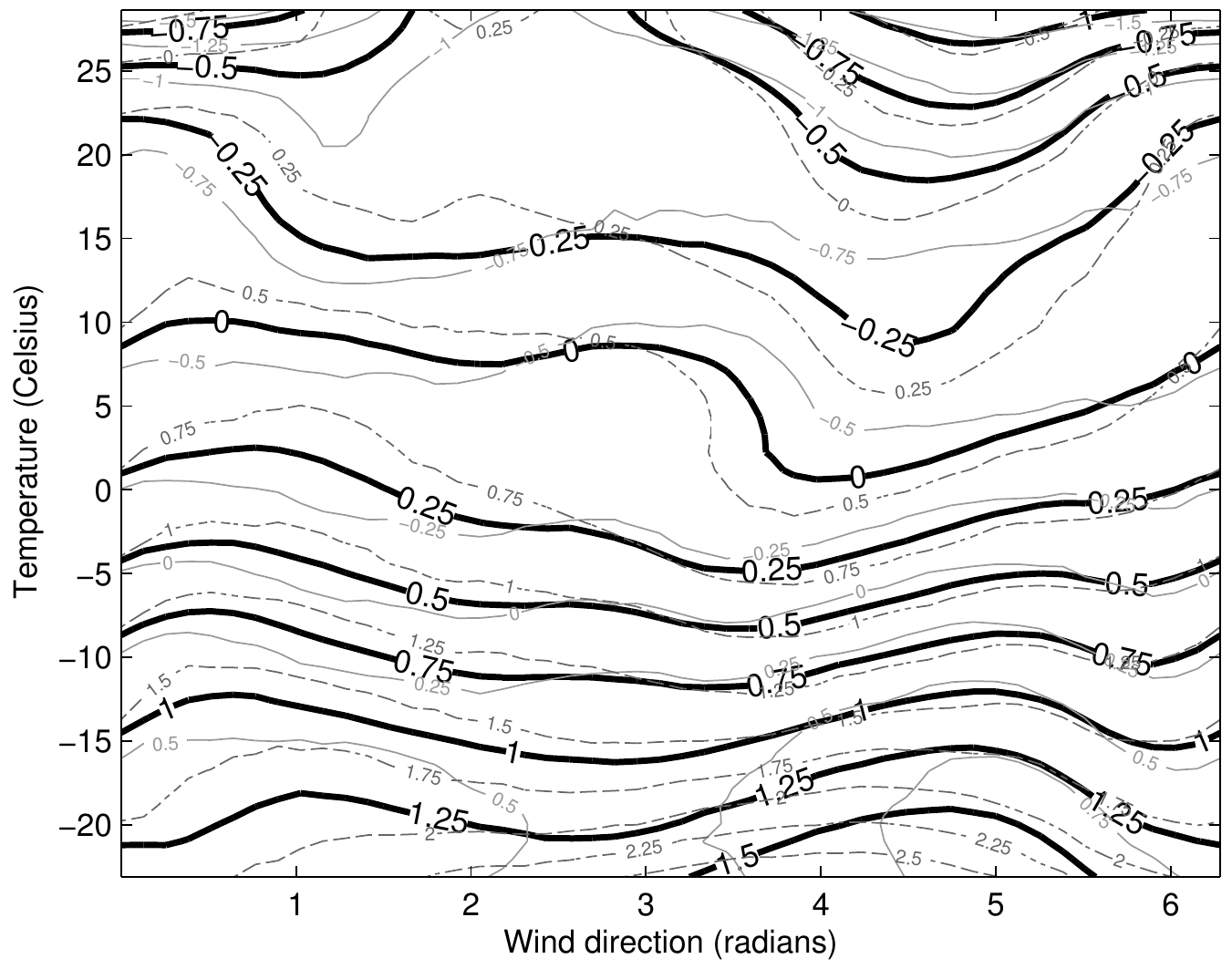}\label{fig:htemp}} 
\subfloat[Traffic]{\includegraphics[width=0.3\textwidth]{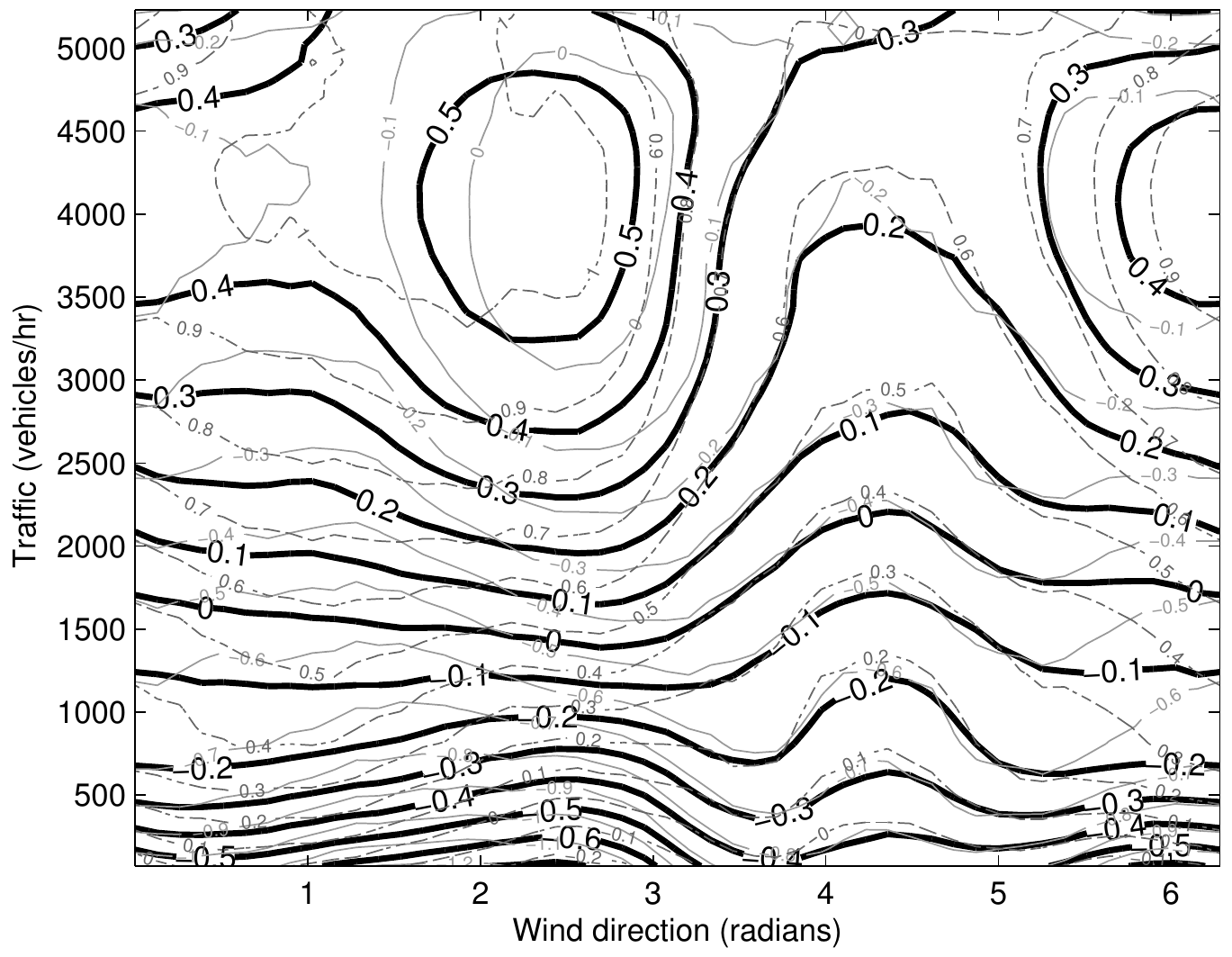}\label{fig:htraffic}} 
\subfloat[Solar radiation]{\includegraphics[width=0.3\textwidth]{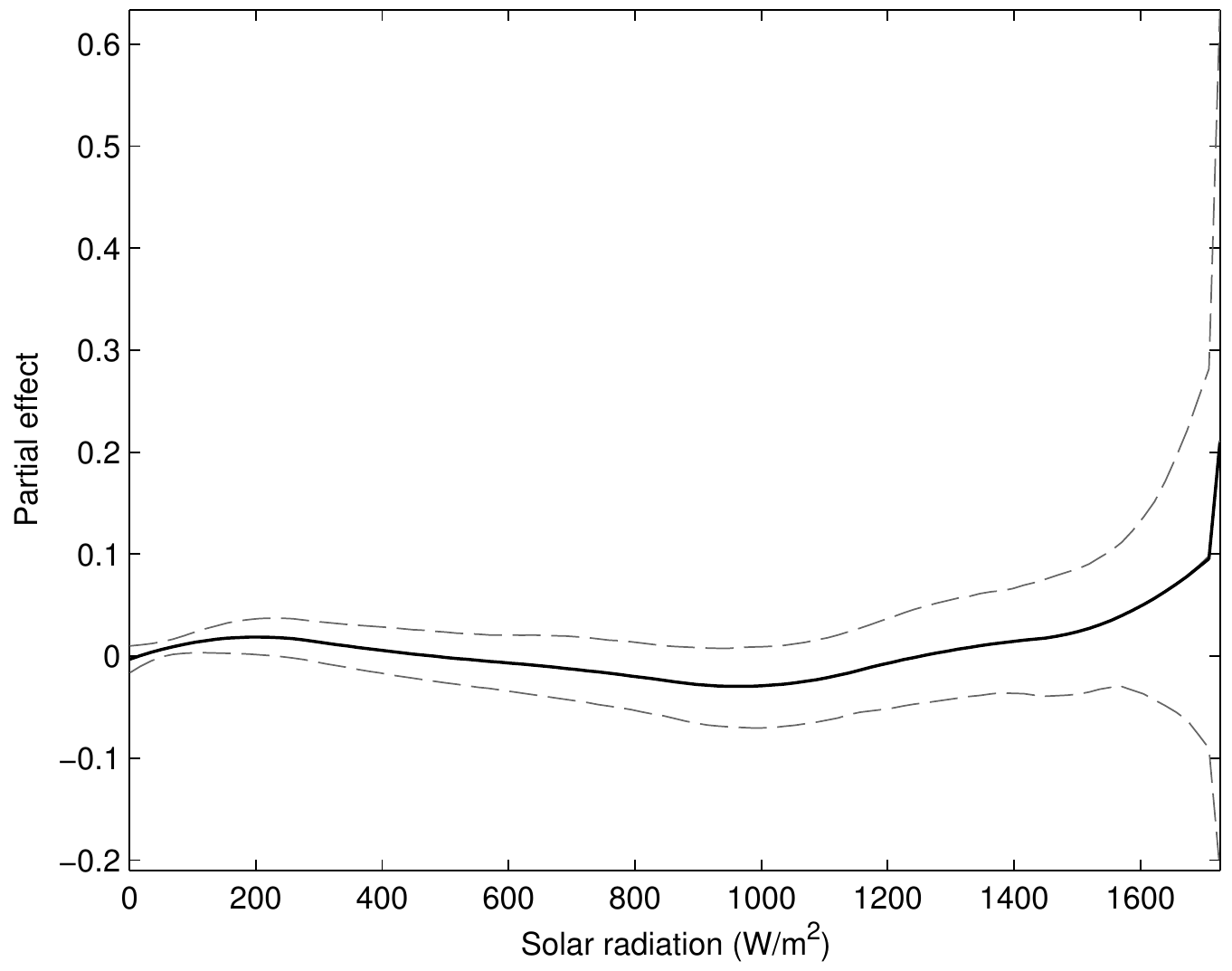}\label{fig:hpar}} 
\caption{Density plots of $\beta_0$, $\sigma$, temporal trends and meteorological covariate effects. 95\% credible intervals are represented as black rugplots for parameters and as dashed lines for fitted smooth terms. The solid grey lines in the contour plots represent the 2.5\% quantiles and the dashed grey lines represent the 97.5\% quantiles.}
\label{fig:finland} 
\end{figure*}

Figure \ref{fig:finland} shows the posterior density estimates and 95\% credible intervals for the mean, standard deviation, \samref{fig:hbeta0} and \samref{fig:hsigma}, and the mean and 2.5\% and 97.5\% quantiles for the fitted smooth functions of temporal trends and non-linear covariate effects \samref{fig:htime} -- \samref{fig:hpar}.

The daily trend in PNC varies over the year (see \samref{fig:htime}) with most days having a peak around 10am. This trend peaks during the summer period (days 140--250) with a daily peak around 11pm and a plateau from approximately 6am to midday. The daily trend in PNC in winter has a trough at 3am and a peak around midday (when most of the day's light occurs). Spring and autumn daily trends contain two peaks, one in the late evening around 9pm and one in the morning aroung 9-10am. The overall shape of these temporal trends is consistent with previously reported temporal trends \citep{molgaard2011}.

The weekly trend \samref{fig:hdow} shows a decreased partial effect on weekends (days 1 and 7) and after accounting for these temporal trends, much of the remaining temporal variation is explained with the Lag 1 autoregressive error although there is still some amount of autocorrelation at Lags 24 and 168, which is captured by the model.

The joint effect of wind speed and wind direction \samref{fig:hwind} is roughly linearly decreasing and almost independent of wind direction for winds weaker than 4m/s. When the wind is blowing above 5 m/s there is an observable non-linear interaction of wind speed and wind direction such that at 11 m/s a wind of angle approximately 170 degrees (3 radians) corresponds to an increase in log PNC and at 30 degrees (0.5 radians) the contribution to log PNC is negative. That is, a strong wind can either remove particles from the microenvironment or transport them from a nearby source.

\begin{figure*}[htbp]
\centering
\subfloat[Marginal daily trend]{\includegraphics[width=0.4\linewidth]{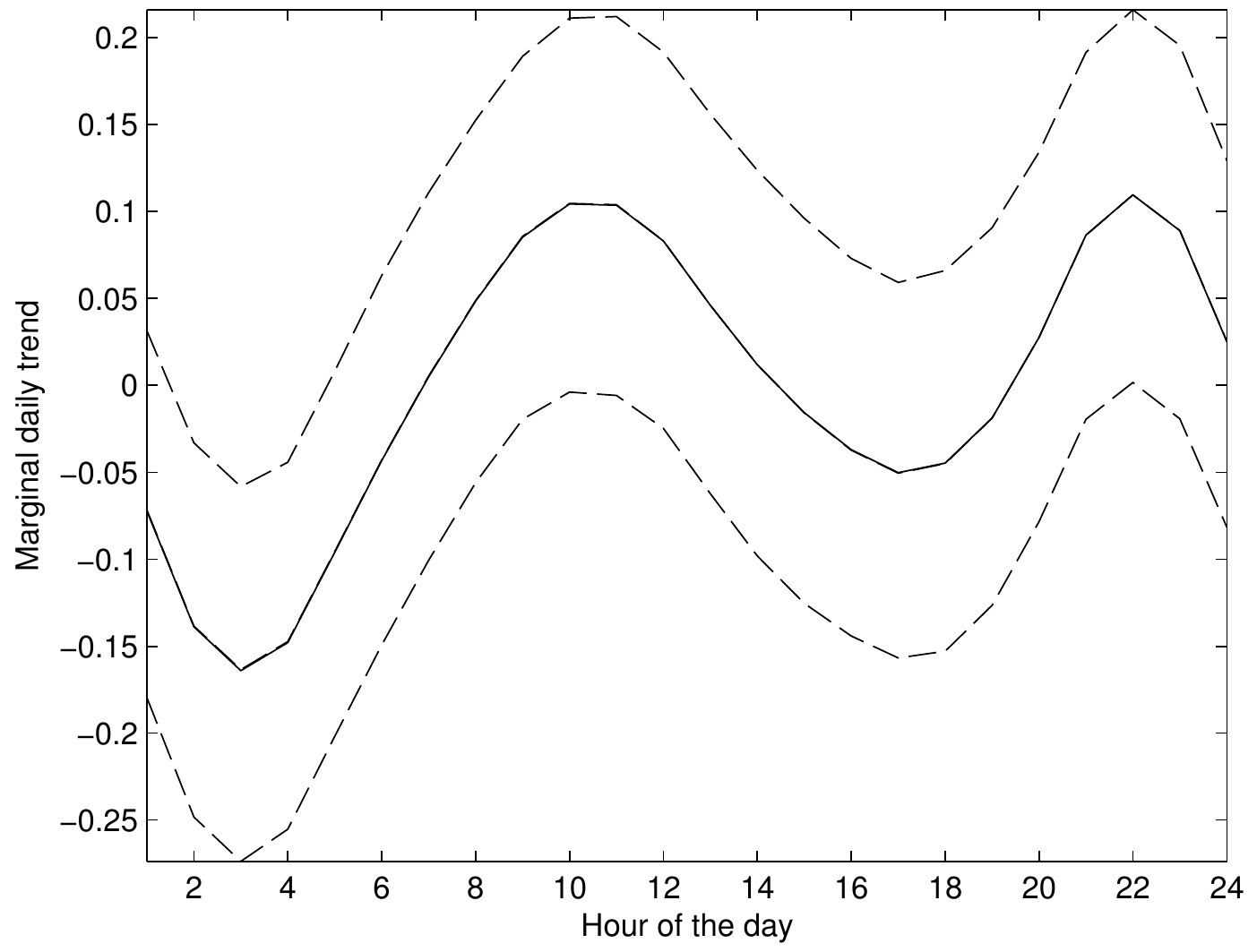}}
\subfloat[Marginal annual trend]{\includegraphics[width=0.4\linewidth]{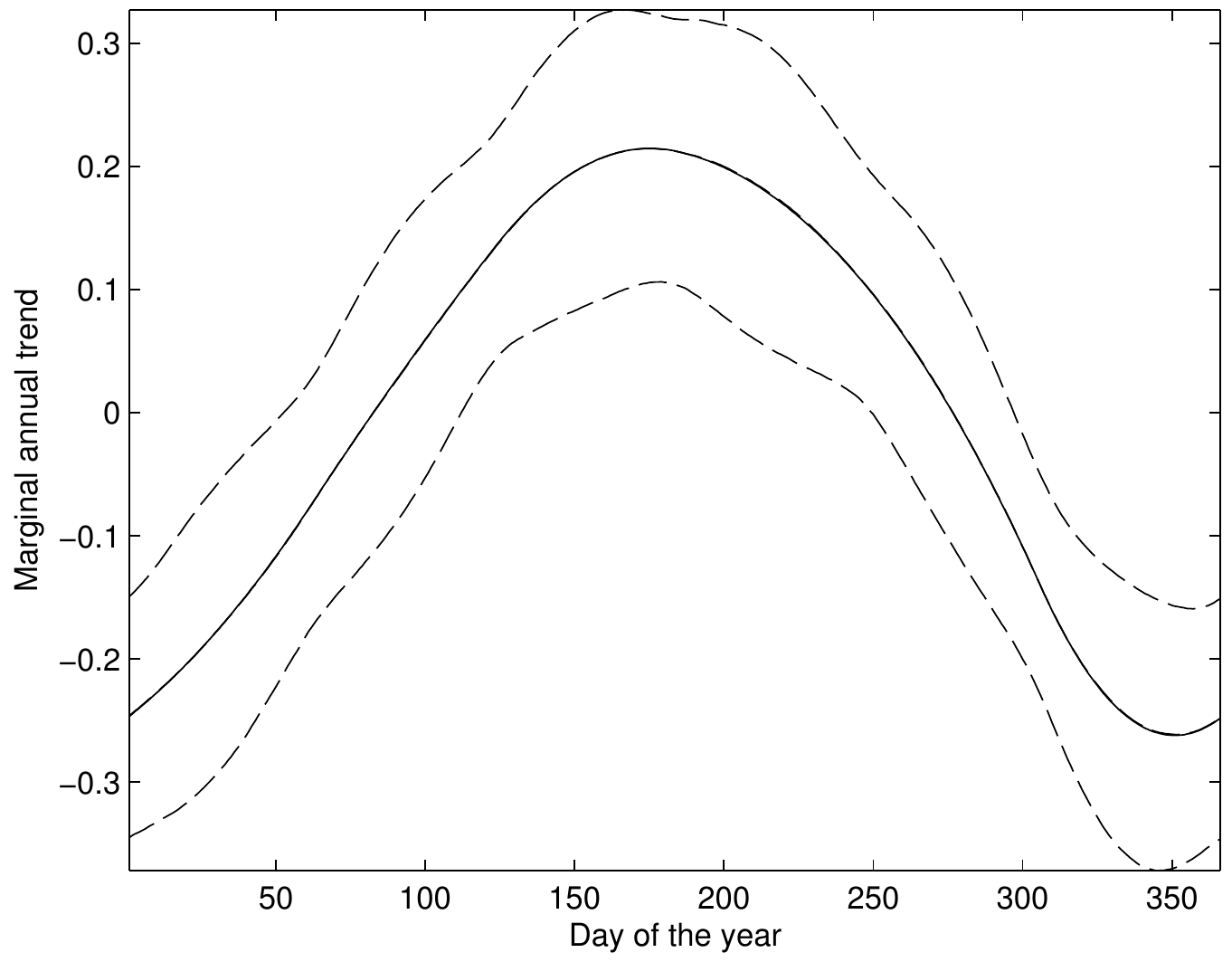}}
\caption{Daily and annual trends calculated by marginalising the posteriors from the joint daily and annual trend.}
\label{fig:martrends}
\end{figure*}

Marginalising this joint trend over the year provides an estimate of the mean daily trend in PNC and vice versa for marginalising over the day to obtain the mean annual trend (Figure \ref{fig:martrends}. The mean daily trend in PNC exhibits two peaks, one at 10am and one at 10pm. The mean annual trend shows a maximum in summer, when the days are long.

Marginalising the posterior for the joint daily and annual trend (Figure \ref{fig:finland}), the mean daily trend in PNC exhibits two peaks, one at 10am and one at 10pm. The mean annual trend shows a maximum in summer, when the days are long (Figure \ref{fig:martrends}).

\begin{figure}[htbp]
\centering
\subfloat[Posterior covariance of $\vec{\varepsilon}$]{\includegraphics[width=0.4\linewidth]{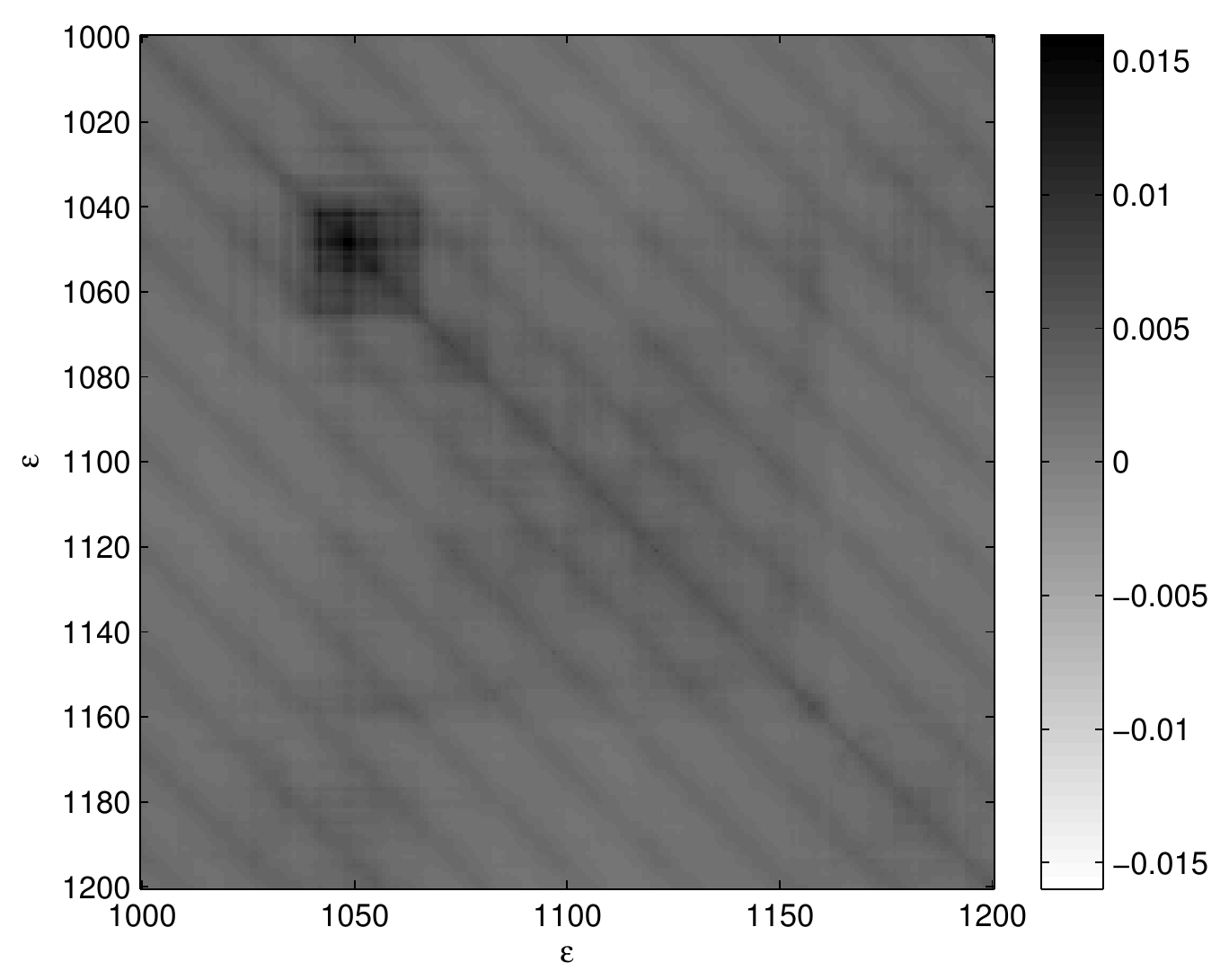}}
\quad
\subfloat[Posterior covariance of $\vec{u}$]{\includegraphics[width=0.4\linewidth]{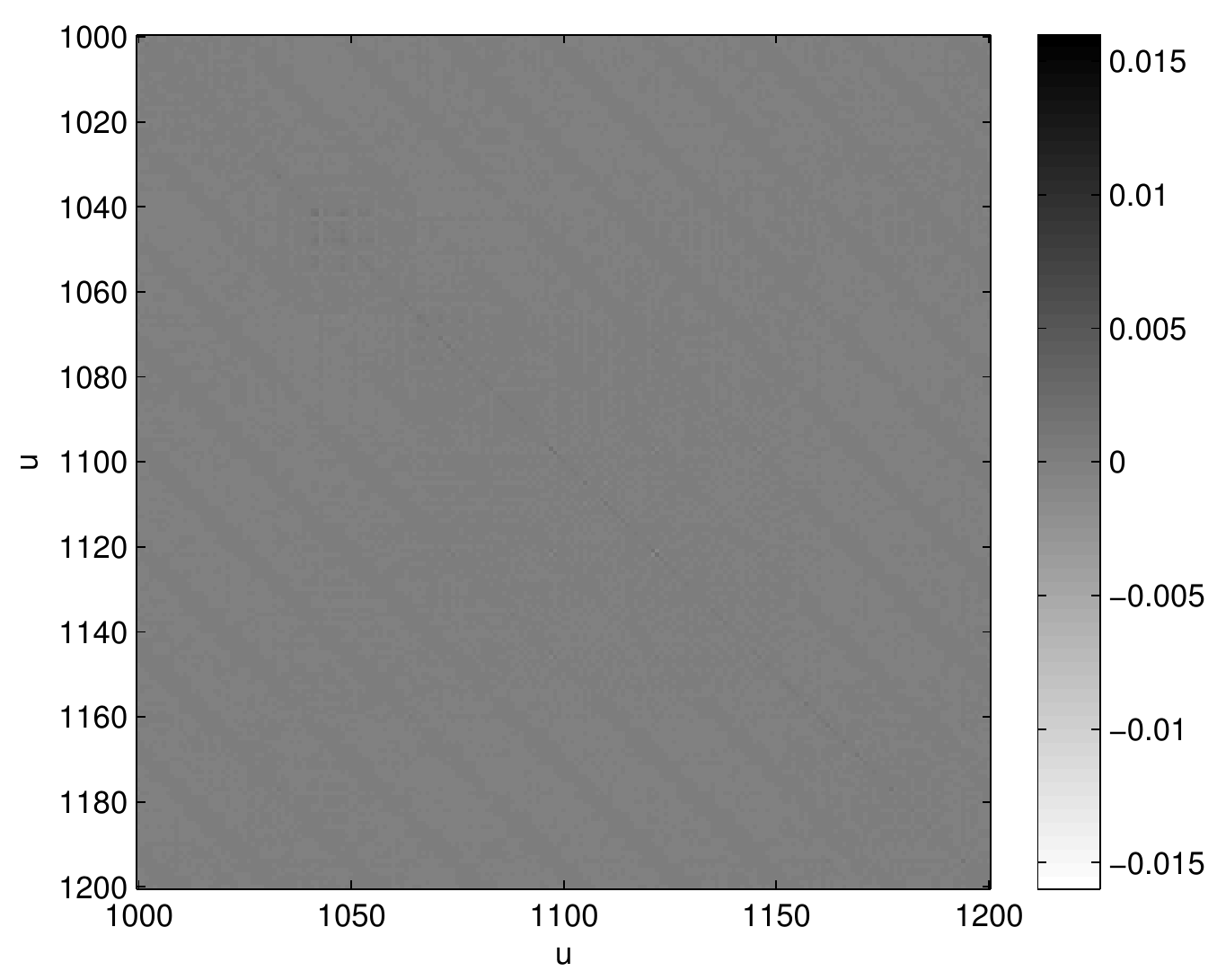}}
\caption{Autocorrelation of residuals estimated from posterior samples of $\vec{\varepsilon}$ and $\vec{u}$.}
\label{fig:helcov}
\end{figure}

The autocovariance of the posterior samples of the residuals for observations 1000 to 1200 are shown in Figure \ref{fig:helcov}. The raw residuals, $\vec{\varepsilon}$, exhibit a noticeable amount of autocorrelation around observation 1050, indicating that while the semi-parametric regression model for the covariates has fitted a smooth annual and daily trend, not all the temporal variability has been explained. This residual variability is captured with by explicitly modelling the autocorrelation in the residuals and we see that the posterior covariance of these residuals decreases significantly when examining $\vec{u}$, the residuals with explicitly modelled autocorrelation.

The effect of high humidity \subref{fig:hrh} above 90\%, is to reduce the total number of particles in the air because of precipitation. The effect in the middle range is fairly flat with some local peaks. The effect of temperature \subref{fig:htemp} is generally decreasing and shows some dependence on wind direction but there are no local peaks or troughs. The effect of traffic density \subref{fig:htraffic} increases steadily up to about 3500 vehicles per hour with some dependence on wind direction. The increase is not as marked at a wind direction of about 230 degrees (4 radians), with a peak occurring around 140 degrees (2.5 radians).

Much of the effect of solar radiation \ref{fig:hpar} is accounted for by the joint daily and annual trend. The estimates of the effect of temperature, wind speed and direction, solar radiation and humidity (as a proxy for rainfall) and weekly trend show very good agreement with those previously reported \citep{Clifford2011}. In both the results presented here and by \citet{Clifford2011} the effects of temperature, humidity and wind speed are generally decreasing. Exceptions include strong wind speeds at certain headings and the positive effect of increased humidity until 90\% as described above.

\begin{figure}[htbp]
\centering
\subfloat[Lag 1]{\includegraphics[width=0.8\linewidth]{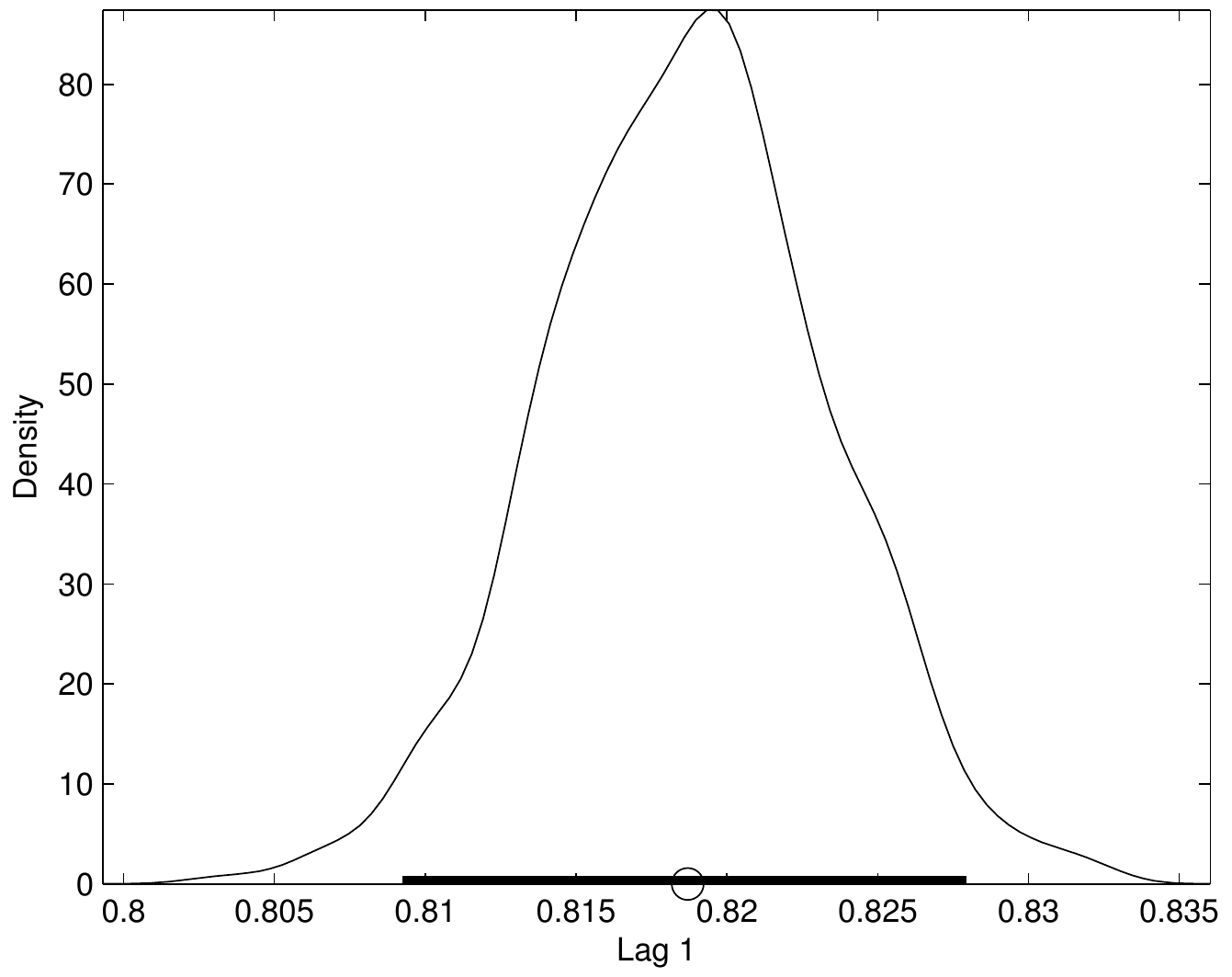}\label{fig:hlag1}} \\
\subfloat[Lag 24]{\includegraphics[width=0.8\linewidth]{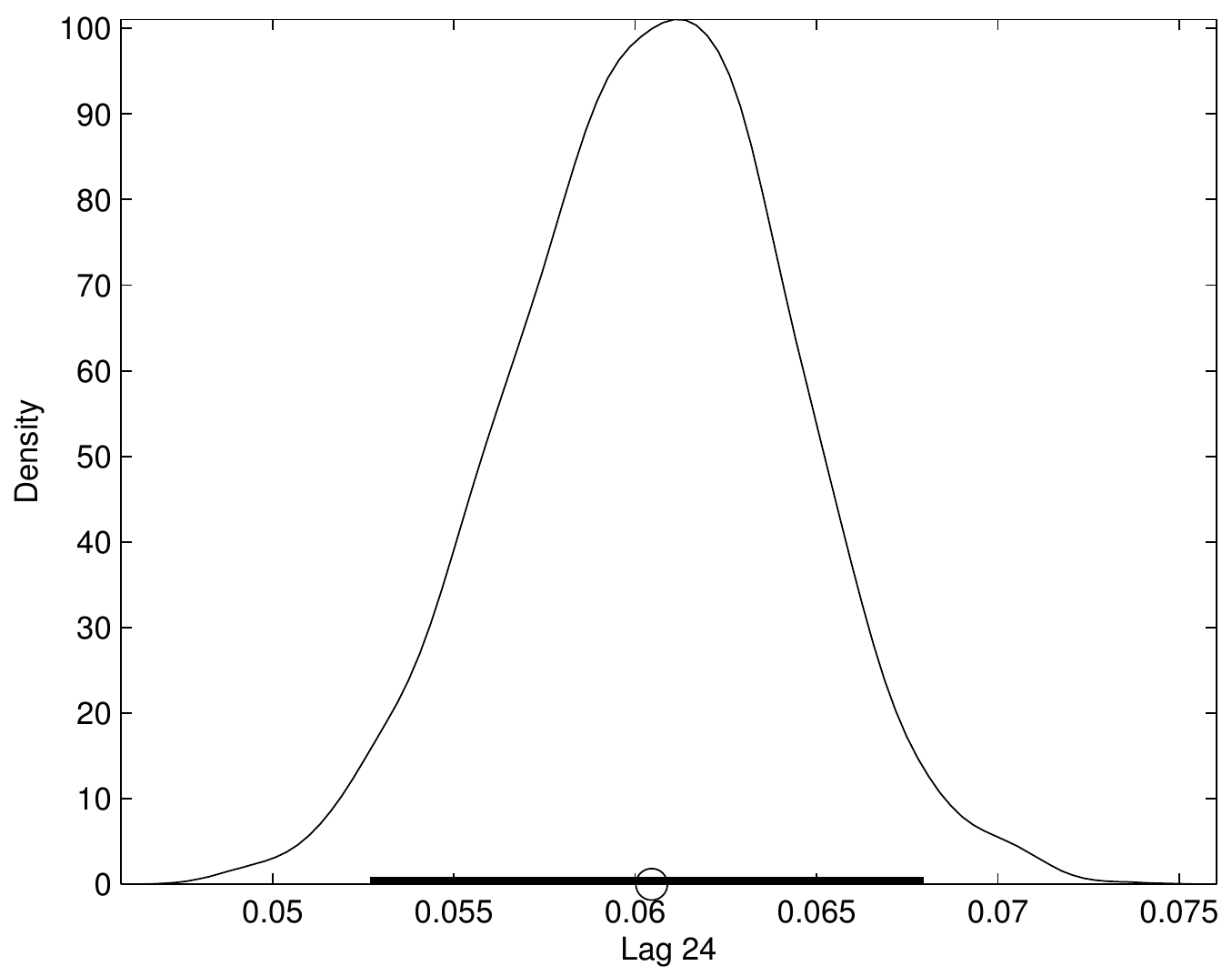}\label{fig:hlag24}} \\ 
\subfloat[Lag 168]{\includegraphics[width=0.8\linewidth]{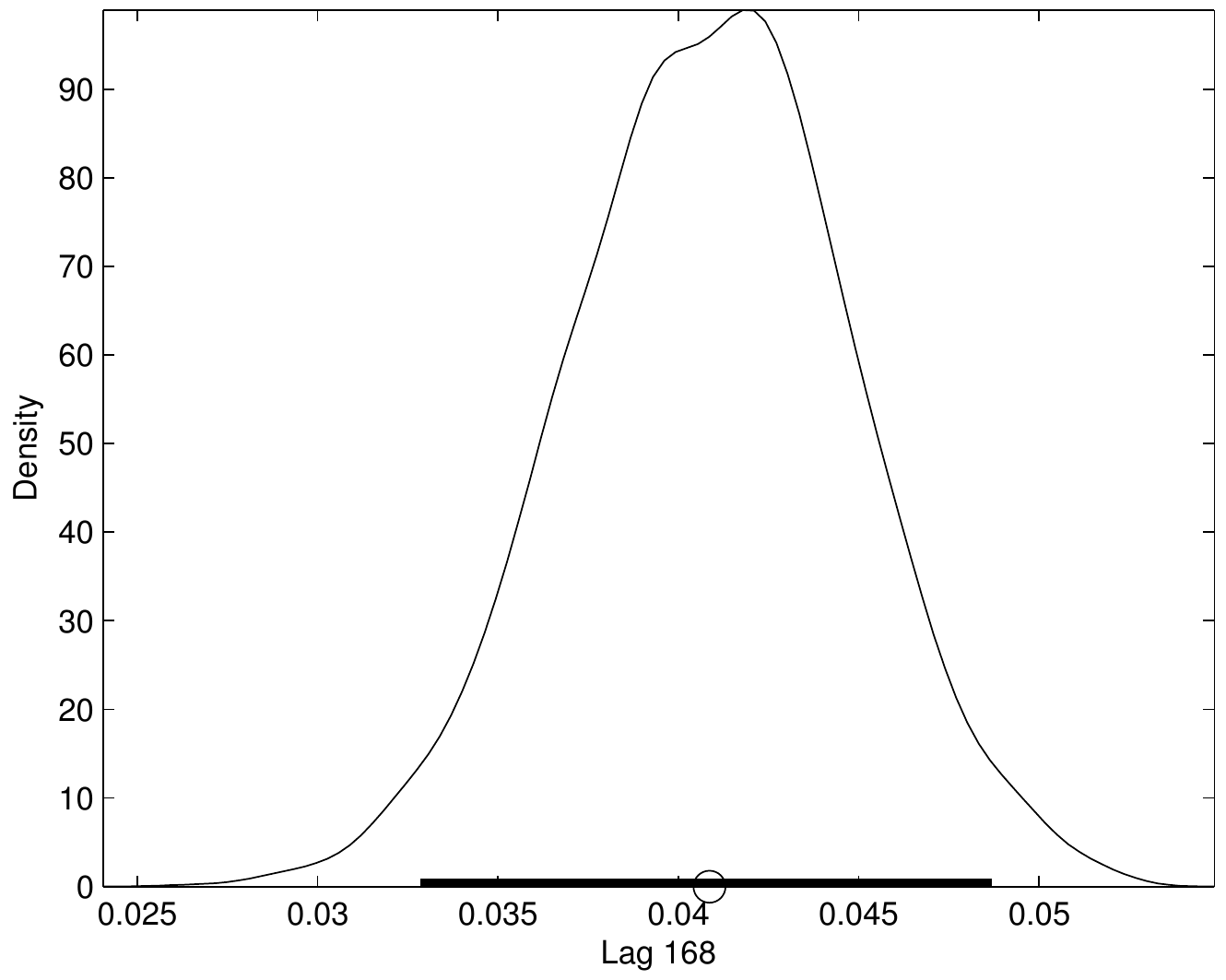}\label{fig:h168}} 
\caption{Density plots of autoregression parameters $\vec{\phi}$. 95\% credible intervals are represented as black rugplots.}
\label{fig:finlandphi} 
\end{figure}

The estimates of $\vec{\phi}$ (Figure \ref{fig:finlandphi}) do not contain zero in their credible intervals. We see a high amount of positive autocorrelation at Lag 1 and a small, but strictly positive, amount at Lags 24 and 168.

By specifying the regression model to include autoregressive residuals rather than autoregressive PNC, the mean daily trend (Figure \ref{fig:martrends}) has peaks which occur at 10am and 10pm rather than at 8am and midday. The temporal variation in the model presented in this section can be expressed as
\begin{multline*}
\log y_i =  f\left( \textnormal{hour}_i, \textnormal{day of the year}_i \right) + \ldots + \\ \left(1 - \phi_1 - \phi_{24} - \phi_{168}\right)^{-1} \varepsilon_i
\end{multline*}
while the temporal variation in the model of \citet{Clifford2011} corresponds to
\begin{multline*}
\log y_i = f_1\left( \textnormal{hour}_i \right) + f_2\left( \textnormal{day of the year}_i \right) + \\ f_3 \left( \log y_{i-1} \right) + f_4 \left( \log y_{i-24} \right) + \ldots + \varepsilon_i.
\end{multline*}
These two specifications are quite different as one contains a joint model of daily and annual trends and models the residuals autoregressively while the other models the residuals as autoregressive. As such, they provide quite different estimates of the daily trend. The weekly trend is the same across both models, with a maximum on Wednesday and minium values on the weekend.

To illustrate the modelling of the autoregressive nature of the residuals a contiguous subset of the modelled values and residuals was randomly selected, corresponding to observations 1000 to 1200. This is approximately eight days of measurements. Analysis of the autocovariance of the posterior samples of the residuals can be found in the appendix.

The effective degrees of freedom for the intercept, each spline term and the model overall are given in Table \ref{tab:finedf}. Note that the intercept has exactly one degree of freedom and that each spline has approximately one less effective degree of freedom than the number of basis splines; this is due to the constraint that each spline is centred around zero. The effective number of parameters for this model according to the DIC is $243.443$. The effective number of degrees of freedom for the linear predictor is $230.643$ (Table \ref{tab:finedf}). These two estimates of model complexity agree quite well given the degrees of freedom given up by the constraint on each spline.

\begin{table*}[htb]
\centering
\begin{tabularx}{\linewidth}{X*{5}{r}}
\toprule
Term & 2.5\%  & 50\%   & 97.5\% & Mean   & SD \\
\midrule
Intercept, $\beta_0$ & 1.00 & 1.00 & 1.00 & 1.00 & 0.00 \\
Wind speed and direction & 46.54 & 47.01 & 47.41 & 47.00 & 0.22 \\
Solar radiation & 6.93 & 6.97 & 6.99 & 6.97 & 0.01 \\
Annual and daily trend & 31.39 & 32.42 & 33.51 & 32.42 & 0.55 \\
Weekly trend & 6.00 & 6.00 & 6.00 & 6.00 & 0.00 \\
Temperature and wind direction & 38.14 & 40.55 & 43.28 & 40.59 & 1.37 \\
Traffic and wind direction & 48.30 & 50.12 & 51.99 & 50.15 & 0.96 \\
Relative humidity and wind direction & 44.22 & 46.44 & 48.92 & 46.52 & 1.18 \\
\midrule
Total for entire model & 227.18 & 230.66 & 233.91 & 230.64 & 1.68 \\
\bottomrule
\end{tabularx}
\caption{Effective degrees of freedom summary statistics}
\label{tab:finedf}
\end{table*}

As described in Section \ref{sec:fore}, estimates of forecast values are obtained by modelling the observed values, modelling the future value based on measured meteorological measurements (in reality, weather forecasts are available instead of observed values) and adding the autocorrelated noise based on the estimates of $\vec{\phi}$. The modelled values (for both the observed and future values) are conditioned on $\vec{\phi}$ and so any inference performed on the forecasting must be done on $\mat{X} \vec{\beta} + \vec{\varepsilon}$ rather than the linear predictor $\mat{X} \vec{\beta}$.

\section{Conclusion}\label{sec:conclusion}
This paper presents a regression method which combines semi-parametric regression, in the form of penalised splines, and a GLM with autoregressive residuals.

It was shown in the simulation case study of section \ref{sec:simfit} that the method described is capable of approximating the underlying smooth functions used to generate the data as well as the autoregressive noise which was added. The resulting smooth reproduced the underlying 2D function without the oscillations which characterise the use of B-splines with many basis functions and a small amount (e.g. none) of smoothing \citep{eilersmarx2010}.

In section \ref{sec:finfit} the modelling methodology was applied to some real data from Helsinki to infer the temporal trends and the effects of various meteorological and physical phenomenon on ultrafine PNC. The resulting fits were consistent with previous studies of ultrafine PNC in Helsinki \citep{Clifford2011, molgaard2011} but provided fitted smooth functions which do not exhibit the oscillations typical of the use of a Fourier series basis.

Despite the models here having a basis size of more than 200 and a number of smoothing parameters and AR parameters, convergence is fast due to the use of the Gibbs sampler and the sum to zero constraint which ensures identifiability. The first 200 samples were discarded for each block of model fitting and the 2000 samples used for posterior inference came from stable chains, yielding normal posterior densities (e.g. Figure \ref{fig:hbeta0}).

By converting the univariate spline bases of the local meteorological covariates (traffic count, temperature, etc.) to a tensor basis of wind direction and the those covariates the DIC was reduced from 13023 to 12404, indicating that the effect of these covariates is dependent on wind direction. This analysis was absent from \citet{Clifford2011} and the move to a penalised B-spline basis improved the flexibility of the model over that described by \citet{molgaard2011}. Fitting a model with the univariate bases and tensor bases for the covariates did not provide a qualitatively different fit and merely inflated the DIC.

The concurrent modelling of the residuals showed a high level of autocorrelation in the residuals at lag 1. The value of the autocorrelation parameters at lags 24 and 168 is small but non-zero and these lags correspond to modelling the variation in the residuals which is left after modelling the daily and weekly trends. Modelling this autocorrelation reduced the magnitude of the posterior covariance of the residuals.

Modelling with penalised splines captured the smooth temporal trends and the autoregressive model for the residuals explained the residual non-smooth variation. By combining these in the same model, rather than doing it in two steps, the MCMC sampler can trade the smooth trend and rough residuals off against each other; the posterior density for $\vec{\beta}$ contains $\mat{X}^{\ast} = \left( 1 - \phi(L) \right) \mat{X}$ and the posterior for $\vec{\phi}$ contains $\mat{E} = \left(\vec{y} - \mat{X}\vec{\beta} \right) \left( 1, L, L^2, \ldots, L^p \right)$. Fitting this model in a two-step process (i.e. modelling the autocorrelation in the residuals \textit{post hoc}) would not allow this trade-off.

With respect to the methodology outlined in Section \ref{sec:meth}, the number of knots for each spline is fixed rather than allowing the number of knots in the splines to vary and using a reversible jump MCMC method \citep{Biller98adaptivebayesian}. The advice of \citet{eilersmarx2010} is to use a large number of knots, perhaps more than is ``necessary'', and to allow the smoothing penalty to control how wiggly the resulting smooth fit is. \citet{ruppertwandcarroll} and \citet{wand2003} suggest choosing the number of knots for a univariate spline as $\min(n/4, 35)$ (for $n$ the number of unique values the covariate takes) to ensure that the non-linear features are captured. The equivalent degrees of freedom for the spline will generally be substantially lower than the number of basis splines used. As such, $p_D$ will typically not increase as more knots are used, because the DIC (and $p_D$) are calculated from the deviance (which stabilises as the basis size is increased).

Modelling with tensor products of splines allows the investigation of the interaction of two covariates which may have non-linear effects and non-linear interactions without specifying a particular functional form, which may often be no better than a subjective guess.

Adding random effects to a GLM converts it to a Generalised Linear Mixed Model. By analogy, random effects can be added to a GAM to form a Generalised Additive Mixed Model (GAMM) \citep{lin99}. A mixed effect spline can be formed by taking the tensor product of a univariate (or higher dimension) spline with an identity matrix representing the different groups for the mixed effect. As the basis vectors stay the same across mixed effect groups, there is no need to include multiple copies of the basis. It is desirable to come up with a way to pass an argument to the model setup which allows the reuse of basis functions, similar to the \texttt{by=} argument in the R package \textsf{mgcv} \citet{mgcv}. This would be a computational improvement over forming the tensor product of a spline and factor term.

The method presented here provides flexible fitting of covariates which may have non-linear effects. The focus has been on temporal trends using cyclic B-spline basis functions with smoothing penalties to ensure there are no spurious oscillations. The anticipated autocorrelation of the residuals has been explicitly modelled to account for much of the temporal variation that remains after removing smooth trends.

\section*{Acknowledgement}
The authors wish to thank the Division of Atmospheric Sciences and Faculty of Science at the University of Helsinki as well as the Institute for Health and Biomedical Innovation and the Discipline of Physics and School of Mathematical Sciences in the Science and Engineering Faculty at Queensland University of Technology, Brisbane, Australia for their support of Sam Clifford in travelling to Helsinki.

The authors also wish to thank Professor Markku Kulmala (University of Helsinki) and Professor Lidia Morawska (QUT) for their feedback and suggestions throughout the development of this work and Professor Tilmann Gneiting for his comments on a previous version of this paper.

\bibliographystyle{model2-names}

\end{document}